\documentclass[a4paper,11pt]{article}
\pdfoutput=1 
\usepackage[english]{babel}
\usepackage[T1]{fontenc}
\usepackage[utf8]{inputenc}	

\usepackage{jheppub}
\usepackage{bm,amssymb,amsmath,amsfonts,slashed,graphicx,multirow,soul,mathtools,xspace,array,comment,color}
\hypersetup{colorlinks=true, allcolors=blue}
\usepackage{adjustbox}
\usepackage{siunitx}
\usepackage{tabularx}
\usepackage{float}
\allowdisplaybreaks
\usepackage{bbold}
\usepackage{subfigure}
\usepackage[normalem]{ulem}
\usepackage{enumerate}
\usepackage{booktabs}
\usepackage{xparse}
\usepackage{etoolbox}
\usepackage{physics}
\usepackage{hypcap}
\usepackage{xcolor}
\usepackage{gensymb}
\usepackage{chngcntr}

\usepackage{hyperref}
\usepackage{fontawesome5}

\newcommand{\ldoublet}{\text{l}}
\newcommand{\qdoublet}{\text{q}}
\newcommand{\usinglet}{\text{u}}
\newcommand{\dsinglet}{\text{d}}
\newcommand{\esinglet}{\text{e}}

\def\be{\begin{equation}}
\def\ee{\end{equation}}

\title{PDF effects in high-mass Drell-Yan SMEFT analyses across flavour space}

\author[1]{David Marzocca,}
\author[1]{Manuel Morales-Alvarado}

\affiliation[1]{INFN, Sezione di Trieste, Via Bonomea 265, 34136, Trieste, Italy}

\emailAdd{david.marzocca@ts.infn.it}
\emailAdd{manuel.morales.alvarado@ts.infn.it}

\abstract{
High-mass Drell-Yan dilepton production provides one of the most sensitive probes of semileptonic four-fermion operators in the Standard Model Effective Field Theory (SMEFT), thanks to the energy growth of the corresponding contributions. 
At the same time, these measurements also constrain parton distribution functions (PDFs) in the large-$x$ region, where PDF uncertainties can mimic or obscure smooth new-physics effects.
In this work we study how the impact of PDF profiling on SMEFT constraints depends on the quark-flavour structure of the effective operators, by performing a joint fit of SMEFT Wilson coefficients and PDF nuisance parameters.
We find that PDF profiling induces a strongly flavour-dependent degradation of the SMEFT sensitivity, both with current data and in HL-LHC projections. The largest broadenings occur for operators involving first-generation quarks, whose effects are correlated with the high-$x$ valence-quark luminosities that dominate the high-mass spectrum. Operators involving heavier quark flavours are less affected, although with non-negligible operator-dependent variations.
As a byproduct of our analysis, we obtain an estimate on the changes induced in the PDF sector by profiling SMEFT effects.
We also show that angular information provides an important handle to reduce degeneracies among SMEFT directions and between SMEFT effects and allowed PDF deformations.
These results demonstrate that the relevance of PDF uncertainties in high-energy SMEFT fits is not uniform across flavour space, and must be assessed in a flavour-dependent way.}

\begin{document}

\maketitle
\flushbottom


\section{Introduction}
\label{sec:intro}

High-energy tails of LHC scattering processes provide powerful probes of heavy new physics. If new particles lie above the partonic energies directly probed, their virtual effects can be described by local operators in an effective field theory. In many relevant cases the interference with the Standard Model (SM) induces corrections that grow with the partonic centre-of-mass energy, $E$, as $E^2/\Lambda^2$, where $\Lambda$ is a typical mass scale of the new heavy particles.

Neutral-current Drell-Yan (DY) dilepton production, $p p \to \ell^+ \ell^-$ with $\ell=e,\mu$, is one of the cleanest channels in which to exploit this strategy. The leptonic final state is experimentally simple, the SM prediction is under good perturbative control, and the high invariant-mass region is directly sensitive to short-distance semileptonic interactions. The Standard Model Effective Field Theory (SMEFT) provides the natural framework to parameterise such effects when no new light degrees of freedom are present. Several studies have shown that high-mass DY data can place constraints on SMEFT operators that are competitive with or complementary to electroweak precision observables and low-energy flavour measurements \cite{Cirigliano:2012ab,deBlas:2013qqa,Farina:2016rws,Greljo:2017vvb,Fuentes-Martin:2020lea,Torre:2020aiz,Crivellin:2021rbf,Crivellin:2021bkd,Allwicher:2022mcg,Allwicher:2022gkm,Greljo:2022jac}.

For similar reasons, the same high-mass DY measurements also constrain parton distribution functions (PDFs), especially in the poorly known large-$x$ region. 
This creates two related but different issues. First, PDF variations and SMEFT effects can both produce smooth distortions of the high-mass dilepton spectra, making it non-trivial to decide whether a deviation should be attributed to new physics or to the proton structure. This effect can be accounted for by properly implementing PDF uncertainties in SMEFT DY studies, as done for instance in Ref.~\cite{Torre:2020aiz}.
The second issue stems from the fact that PDF sets used as input to SMEFT interpretations may already contain information from the data being interpreted, or data that is affected by the same SMEFT operators being considered, such as other high-mass DY datasets.
A statistically consistent way to address both ambiguities is to include PDF and SMEFT degrees of freedom in the same likelihood from the start, as done in Refs.~\cite{Carrazza:2019sec,Greljo:2021kvv,CMS:2021yzl,Iranipour:2022iak,Gao:2022srd,Kassabov:2023hbm,Hammou:2023heg,Costantini:2024xae,Cole:2026eex}.
These studies showed that profiling over PDF degrees of freedom can significantly weaken SMEFT constraints relative to fixed-PDF analyses, an effect already visible with current data and expected to become more important at the HL-LHC.

These previous works have addressed the interplay between PDFs and SMEFT effects either in selected benchmark scenarios or within full simultaneous PDF--SMEFT fitting frameworks. In this work we focus instead on a complementary question: how the impact of PDF profiling depends on the quark-flavour structure of the effective operators. Such a dependence is expected, since the relevant PDF uncertainties are themselves strongly flavour dependent, with different behaviours for valence quarks, light sea quarks, and heavier flavours.

To tackle the two issues mentioned above, we adopt a simplified two-stage strategy. We start with a PDF set constructed from a fit that excludes high-mass DY datasets, and encode PDF uncertainties in a Hessian PDF set.
In a Bayesian approach, we then combine this prior likelihood for the PDF parameters with the high-mass DY likelihood, including both the PDF and SMEFT dependence of the latter. 
In this way, we can implement PDF uncertainties in SMEFT studies while avoiding using high-mass DY data for the prior PDF likelihood.
This approach also avoids repeating a full global PDF fit whenever new DY data are added. The limitation is that a fixed Hessian prior can only describe deformations within the uncertainty space of the prior fit; it cannot capture large non-linear changes, non-Gaussian effects, or substantial rearrangements of the PDF uncertainty directions induced by the new data. The method is therefore appropriate only when the added high-mass DY data constrain the prior PDFs without dominating their determination. We will verify a posteriori that this condition is satisfied in the scenarios considered here.

The paper is structured as follows. In Section~\ref{sec:exp} we present the experimental analysis for high-mass DY that we employ. In Section~\ref{sec:theory} we discuss the theoretical framework of the analysis regarding SMEFT and PDF aspects, as well as the statistical analysis. Our results are collected and discussed in Section~\ref{sec:results}. 
The one-parameter 95\% CL intervals are reported numerically in the public results repository
\textcolor{black}{\faGithub}\ 
\href{https://github.com/manuel-morales-a/flavourful-drell-yan-smeft-results}
{\texttt{flavourful-drell-yan-smeft-results}}. 
In the same repository we also publish a Mathematica notebook with an analytic approximation of the posterior $\chi^2$, profiled over all nuisance parameters, as a quartic polynomial of the SMEFT coefficients.
We conclude in Section~\ref{sec:conclusions}.
Further technical details and supplementary results are collected in Appendices~\ref{app:pdf}, \ref{app:xsec}, and \ref{app:fl_general}.

\section{Experimental data}
\label{sec:exp}

High-mass dilepton production has been studied extensively at the LHC, both for Standard Model measurements \cite{ATLAS:2013xny,CMS:2013zfg,CMS:2014jea,ATLAS:2016gic,ATLAS:2017rue,CMS:2018mdl} and searches for new physics in dilepton tails \cite{ATLAS:2019erb,ATLAS:2020yat,CMS:2021ctt,ATLAS:2021mla,CMS:2025tlo}.

A maximally informative analysis would combine all relevant ATLAS and CMS measurements at different centre-of-mass energies. In this work, however, our aim is a first exploration of a simultaneous SMEFT-PDF fit in a generic quark-flavour scenario, rather than a complete experimental combination. We therefore restrict the analysis to a single public dataset with high sensitivity to contact interactions, large-$x$ PDFs, and with sufficient information to construct a reinterpretation likelihood.

We choose the CMS search for resonant and non-resonant new phenomena in high-mass dilepton final states at $\sqrt{s}=13~{\rm TeV}$, based on the full Run-II dataset collected between 2016 and 2018~\cite{CMS:2021ctt}. The dataset corresponds to an integrated luminosity of $137~{\rm fb}^{-1}$ in the dielectron channel and $140~{\rm fb}^{-1}$ in the dimuon channel. We use the public HEPData release associated with the non-resonant search, in particular the supplementary inputs provided for the contact-interaction limit calculation.

The relevant observable is the two-dimensional binned distribution in the dilepton invariant mass, $m_{\ell\ell}$, and the scattering angle $\cos\theta^*$, restricted to the two angular regions $\cos\theta^* < 0$ and $\cos\theta^* \geq 0$.\footnote{The angle $\theta^*$ is defined in the Collins-Soper frame. At leading order, where the dilepton transverse momentum vanishes, it is directly related to the polar scattering angle in the partonic centre-of-mass frame, up to the convention used to infer the incoming quark direction from the longitudinal boost of the dilepton system. The precise definition used in this work is given in App.~\ref{app:xsec}.} This binning is well suited for SMEFT interpretations of high-energy Drell-Yan. The invariant-mass dependence captures the energy growth of contact interactions, while the angular information helps distinguish different helicity structures, in particular same- and opposite-chirality quark-lepton currents.

The CMS release provides event yields, SM background estimates, as well as a full covariance matrix for systematic uncertainties, in exclusive bins of \(m_{\ell\ell}\) and \(\cos\theta^*\), separately for the electron and muon final states, split by data-taking year and detector category. We aggregate these inputs over the three Run-II periods and over the lepton pseudorapidity categories, while retaining the bin-to-bin systematics covariance information provided in the HEPData release.\footnote{If $A$ is the aggregation matrix, with entries equal to one when a fine bin contributes to a given coarse bin and zero otherwise, the aggregated yields and covariance matrix are
\begin{equation*}
  N_{\rm obs} = A N_{\rm obs}^{\rm fine},
  \qquad
  N_{\rm SM} = A N_{\rm SM}^{\rm fine},
  \qquad
  V = A V^{\rm fine} A^T ~,
\end{equation*}
where $N_{\rm obs}$ denotes the observed yields, $N_{\rm SM}$ the central SM background prediction, and $V$ the correlated experimental covariance matrix in yield space.}
This gives a common binned representation for each lepton channel, equivalent to the mass-angle spectra shown in Figure~3 of Ref.~\cite{CMS:2021ctt}, but with the correlated experimental uncertainties propagated consistently to the fit.
We include in the fit only bins with dilepton invariant mass larger than 400~GeV, since lower energy bins are possibly already included in the dataset used to obtain the PDF set we employ and because the SMEFT effects studied here become large only at high energies. The total number of aggregated bins amounts to 12 for dielectron data and 12 for dimuon data.

\begin{figure}[t]
    \centering
    \includegraphics[width=0.95\linewidth]{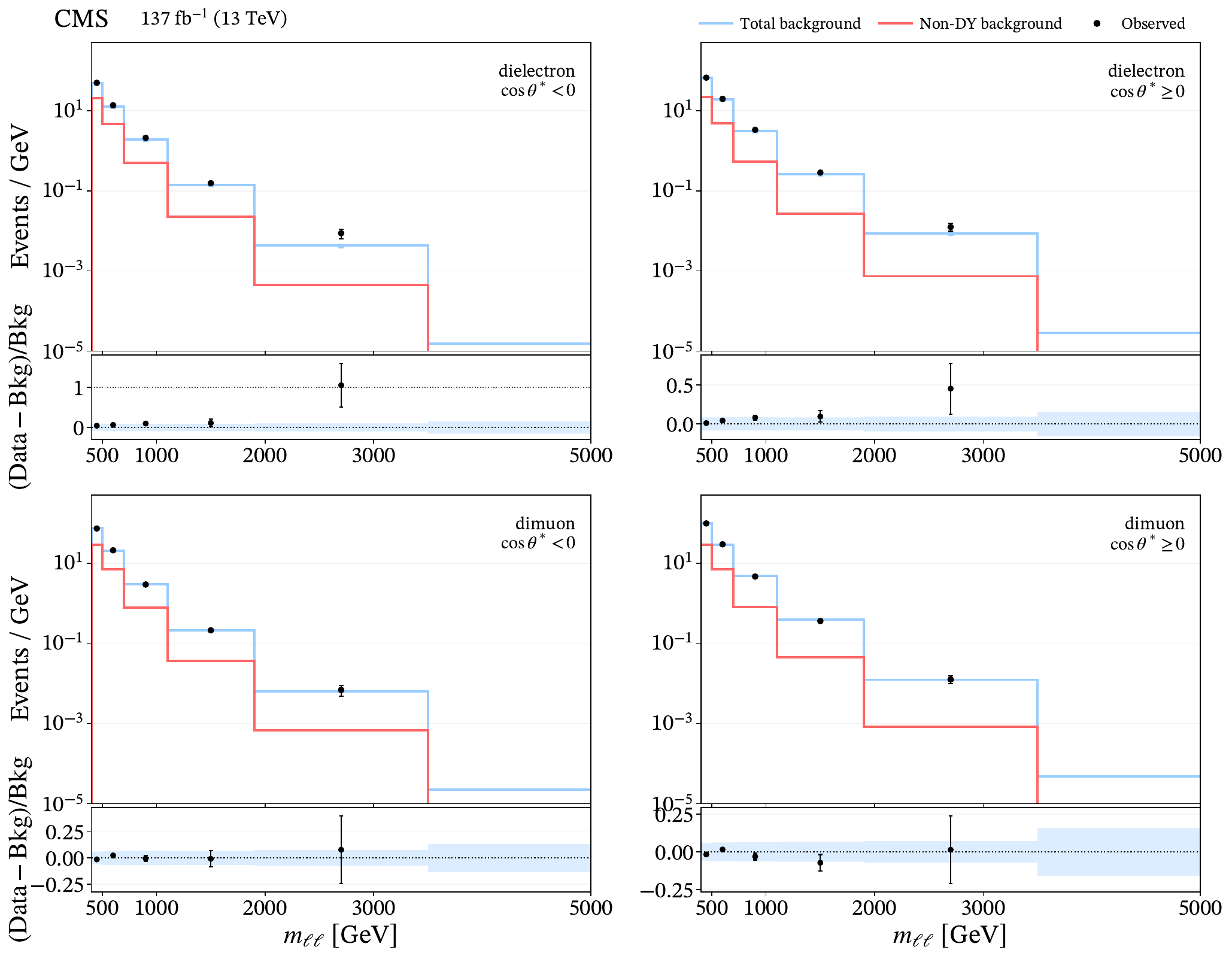}
    \caption{Observed event counts and expected background yields in the CMS 2021 high-mass dilepton contact-interaction bins used in this analysis. The DY and non-DY components are obtained from our approximate background split, normalised to the official CMS total background in each bin. In the ratio plots we also show the CMS systematic uncertainty on the expected event count for each bin.}
    \label{fig:CMS2021data}
\end{figure}

The CMS analysis provides the total expected SM background in each bin. For the purposes of separating the SMEFT-sensitive DY contribution from the remaining backgrounds, we obtain an approximate DY/non-DY decomposition from the stacked-background information in Figure~3 of Ref.~\cite{CMS:2021ctt} and normalise the sum to the official CMS total background. Thus, for each bin $I$ defined by the lepton flavour $\ell_k$ and the $m_{\ell\ell}$ and $\cos\theta^\ast$ intervals, we obtain the central SM prediction as
\begin{equation}
  N^{\rm SM,CMS}_{I} = N^{\rm DY,CMS}_{I} + N^{\text{non-DY,CMS}}_{I}\, .
  \label{eq:total_sm}
\end{equation}
The expected and observed event counts in each bin are shown in Figure~\ref{fig:CMS2021data}.
The non-DY contribution is sizeable in the lowest mass bins, but becomes subleading in the high-mass tail, where it is typically several times smaller than the DY contribution and, above about \(2~{\rm TeV}\), approximately an order of magnitude smaller.

\subsection{High luminosity projections}
\label{subsec:hl_data}

We also explore high-luminosity projections of the same high-mass Drell-Yan observables. 
The purpose of our projections is not to reproduce a future experimental analysis in full detector-level detail, but to estimate how the SMEFT sensitivity of the present measurement would evolve in an HL-LHC scenario. 
Since the larger integrated luminosity will reduce statistical uncertainties, correlated systematic uncertainties are expected to become increasingly important. We therefore use a simplified projection that preserves the binning and the correlated systematic structure of the CMS analysis in Ref.~\cite{CMS:2021ctt}.
The drawback of this approach is the fact that we use the same wide $m_{\ell\ell}$ and $\cos \theta^*$ bins for HL-LHC as in the CMS analysis, while a finer binning would be possible at higher luminosity.

In practice, we obtain the SM expectation in each bin by rescaling the current CMS expected event yields. For the Drell-Yan component we define the rescaling ratios
\begin{equation}
  K^{\rm DY,HL}_{I}
  =
  \frac{\mathcal{L}_{\rm HL}}{\mathcal{L}^{\ell}_{\rm CMS}}\,
  \frac{\sigma^{\rm SM-DY}_{I}(14~{\rm TeV})}
       {\sigma^{\rm SM-DY}_{I}(13~{\rm TeV})}\, ,
  \label{eq:hl_dy_rescaling}
\end{equation}
where $I$ labels the $(m_{\ell\ell},\cos\theta^*)$ bin and the lepton channel, and $\mathcal{L}^{\ell}_{\rm CMS}$ is the Run-II luminosity of the corresponding CMS channel. 
The ratio of cross sections accounts for the change in centre-of-mass energy from $13$ to $14~{\rm TeV}$, using the same bin definitions as in the Run-II analysis.
The cross section increase ranges from about 11\% in the lowest energy bins to 30\% in the $[1.9,3.5]$~TeV bins and $50$--$60\%$ in the $[3.5,10]$~TeV ones.
For the HL-LHC integrated luminosity we take $\mathcal{L}_{\rm HL}=6\,{\rm ab}^{-1}$, corresponding to an effective ATLAS+CMS combination.

For the subleading non-DY backgrounds we use the same rescaling factor,
\begin{equation}
  K^{\rm non-DY,HL}_{I}
  \simeq
  K^{\rm DY,HL}_{I}
  \equiv K^{\rm HL}_{I}\, .
\end{equation}
This is an approximation since top, diboson, and other reducible backgrounds do not have the same energy dependence as dilepton production. The impact of this approximation is however small in the high-mass bins that dominate the SMEFT sensitivity, where the non-DY contribution is one order of magnitude smaller than the DY one. The projected SM expectation is therefore
\begin{equation}
  N^{\rm SM,HL}_{I} =
  K^{\rm HL}_{I}\left(N^{\rm DY,CMS}_{I} 
  + N^{\text{non-DY,CMS}}_{I} \right) \, .
  \label{eq:total_sm_hl}
\end{equation}
The projected observed counts are constructed with an Asimov prescription~\cite{Cowan:2010js}, $N^{\rm obs,HL}_{I}=N^{\rm SM,HL}_{I}$.
The pseudodata therefore contain no statistical fluctuations around the SM prediction, and the resulting bounds should be interpreted as expected sensitivities under the SM hypothesis.

The correlated experimental systematics are extrapolated from the Run-II covariance matrix by preserving the Run-II fractional covariance between bins. If $V^{\rm Run-II}_{IJ}$ denotes the covariance matrix for the Run-II background prediction, we take
\begin{equation}
  V^{\rm HL}_{IJ} =
  r_{\rm syst}^{\,2}\,
  V^{\rm Run-II}_{IJ}\,
  K^{\rm HL}_{I}\, K^{\rm HL}_{J}\, .
  \label{eq:hl_cov}
\end{equation}
The parameter $r_{\rm syst}$ controls the assumed improvement in systematic uncertainties. The choice $r_{\rm syst}=1$ keeps the Run-II fractional systematic uncertainties unchanged, while $r_{\rm syst}<1$ models improved detector calibration, reconstruction, and background modelling in the HL-LHC phase. 
We use $r_{\rm syst}=0.2$ as our baseline, as in \cite{Greljo:2021kvv}, and also discuss the difference with the more conservative benchmark $r_{\rm syst}=1$ in Section~\ref{subsec:lfu_scenario}.

\section{Theoretical framework}
\label{sec:theory}

The starting point for our prediction of the event yield in each high-mass Drell-Yan bin is the SM background estimate provided by CMS \cite{CMS:2021ctt}, see Eqs.~(\ref{eq:total_sm},\ref{eq:total_sm_hl}).
In particular, the CMS prediction for the Drell-Yan component includes perturbative QCD corrections up to NNLO and electroweak corrections at NLO.
In this section we describe how we supplement this baseline with the dependence on SMEFT coefficients and on PDF nuisance parameters.

\subsection{SMEFT contributions to high-mass Drell-Yan}
\label{sec:SMEFT}

We describe possible short-distance new-physics effects in high-mass dilepton production using the SMEFT.
In this framework, heavy degrees of freedom above the experimentally accessible scale, $\Lambda \gg E$, are integrated out and their leading effects are encoded in higher-dimensional operators built from SM fields.
If lepton number is conserved, the leading corrections arise from dimension-six operators. We work in the Warsaw basis \cite{Grzadkowski:2010es}.

Only a subset of SMEFT operators can give sizeable effects in the high-mass Drell-Yan tails while evading existing constraints from electroweak precision data and low-energy measurements.
We therefore impose two selection criteria, following the logic of Ref.~\cite{Greljo:2017vvb}.
First, we keep operators that induce tree-level corrections to the SM partonic cross section growing at least as $E^2/\Lambda^2$. This selects semileptonic four-fermion operators.
Second, we require a non-vanishing interference with the SM Drell-Yan amplitude in the massless-fermion limit, so that the leading $E^2/\Lambda^2$ correction is present. This excludes scalar and tensor semileptonic operators, as well as flavour off-diagonal neutral currents. 
The excluded directions are not necessarily uninteresting.
While rare leptonic and semileptonic decays of kaons and $B$ mesons currently test down-quark flavour-changing operators at scales much higher than those reachable from high-mass DY at HL-LHC, see e.g. Table~4 of Ref.~\cite{Glioti:2025zpn}, flavour violation in charm is less well tested at low energy and it has been shown that LHC bounds are already competitive \cite{Fuentes-Martin:2020lea}. Also, scalar and tensor operators can arise in motivated UV scenarios, for instance from scalar leptoquark exchange.
For this reason, we plan to generalise the analysis by removing the requirement of a non-vanishing interference with the SM in a future work, focusing now on studying the impact of PDF profiling on the flavour-conserving coefficients with different quark flavours.

The two conditions discussed above limit the SMEFT Lagrangian relevant for this analysis to
\begin{equation}
\begin{aligned}
\mathcal L_{\rm SMEFT}\supset
& \,
C^{(3)[kkii]}_{lq}
(\bar \qdoublet_i\gamma_\mu\sigma^a \qdoublet_i)(\bar \ldoublet_k\gamma^\mu\sigma^a \ldoublet_k)
+
C^{(1)[kkii]}_{lq}
(\bar \qdoublet_i\gamma_\mu \qdoublet_i)(\bar \ldoublet_k\gamma^\mu \ldoublet_k)
\\
&+
C^{[kkii]}_{eu}
(\bar \usinglet_i\gamma_\mu \usinglet_i)(\bar \esinglet_k\gamma^\mu \esinglet_k)
+
C^{[kkii]}_{ed}
(\bar \dsinglet_i\gamma_\mu \dsinglet_i)(\bar \esinglet_k\gamma^\mu \esinglet_k)
\\
&+
C^{[kkii]}_{lu}
(\bar \usinglet_i\gamma_\mu \usinglet_i)(\bar \ldoublet_k\gamma^\mu \ldoublet_k)
+
C^{[kkii]}_{ld}
(\bar \dsinglet_i\gamma_\mu \dsinglet_i)(\bar \ldoublet_k\gamma^\mu \ldoublet_k)
\\
&+
C^{[iikk]}_{qe}
(\bar \qdoublet_i\gamma_\mu \qdoublet_i)(\bar \esinglet_k\gamma^\mu \esinglet_k) ~ .
\end{aligned}
\label{eq:full_basis}
\end{equation}
Here, $\qdoublet_i$ and $\ldoublet_k$ denote the left-handed quark and lepton doublets, while $\usinglet_i$, $\dsinglet_i$, and $\esinglet_k$ are right-handed singlets. The indices $i,k$ label fermion generations, and $\sigma^a$ are the Pauli matrices acting on $SU(2)_L$ indices. CKM mixing inside the quark doublets is neglected.
The Wilson coefficients $C$ have mass dimension $-2$ and in the following we express their values in units of TeV$^{-2}$, absorbing the usual $1/\Lambda^2$ suppression due to the heavy new physics scale into the coefficient.
Excluding the operators involving the top quark, which do not contribute to $pp\to\ell^+\ell^-$, Eq.~\eqref{eq:full_basis} gives 18 independent four-fermion operators for each charged-lepton flavour.

It is useful to describe the effect of these operators directly at the level of the partonic amplitude.
This makes explicit how the SMEFT coefficients modify the short-distance hard scattering, while PDFs enter through the convolution with the initial-state parton luminosities.
Following the parametrisation of Ref.~\cite{Greljo:2017vvb}, the amplitude for $q_i\bar q_i\to \ell_k^-\ell_k^+$, where $q_i \in \{ u_i,d_i \}$ and $\ell_k = (e, \mu)_k$, can be written as
\begin{equation}
\mathcal A(q_i\bar q_i\to \ell_k^-\ell_k^+)
=
i\sum_{\chi,\eta=L,R}
(\bar q_{\chi i}\gamma_\mu q_{\chi i})
(\bar \ell_{\eta k}\gamma^\mu \ell_{\eta k})\,
F_{q_\chi \ell_\eta}^{ik}(p^2)~,
\label{eq:dy_amp}
\end{equation}
in which $p$ is the momentum flowing through the neutral-current interaction.
The form factor $F_{q_\chi \ell_\eta}^{ik}(p^2)$ contains the SM photon and $Z$ exchange contributions, together with the local contact interaction induced by the semileptonic operators:
\begin{equation}
F_{q_\chi \ell_\eta}^{ik}(p^2)
=
\frac{e^2 Q_q Q_\ell}{p^2}
+
\frac{g_Z^{q_\chi} g_Z^{\ell_\eta}}
     {p^2-m_Z^2+i m_Z\Gamma_Z}
+
\epsilon_{q_\chi \ell_\eta}^{ik}~.
\label{eq:dy_ff}
\end{equation}
Here $Q_q$ and $Q_\ell$ are the electric charges, while $g_Z^{f_\chi}$ denotes the chiral SM coupling to the $Z$ boson, normalised as $g_Z^{f_\chi}=2m_Z/v \, (T^3_{f_\chi}-Q_f\sin^2\theta_W)$.
The contact terms are related to the SMEFT coefficients in Eq.~\eqref{eq:full_basis} by
\begin{equation}
    \begin{split}
        & \epsilon_{u_L e_L} = C_{lq}^{(-)} \equiv C_{lq}^{(1)} - C_{lq}^{(3)}\,, \quad
        \epsilon_{d_L e_L} = C_{lq}^{(+)} \equiv C_{lq}^{(1)} + C_{lq}^{(3)}\,, \quad
        \epsilon_{u_R e_L} = C_{lu} \,, \\
        & 
        \epsilon_{u_R e_R} = C_{eu}\, , \quad
        \epsilon_{d_R e_R} = C_{ed}\, , \quad
        \epsilon_{d_R e_L} = C_{ld} \,, \quad
        \epsilon_{u_L e_R} = \epsilon_{d_L e_R} = C_{qe} \,,
    \end{split}
    \label{eq:contact_interactions}
\end{equation}
where we suppressed flavour indices.
The last equality is the only unavoidable $SU(2)_L$ relation among the charged-lepton neutral-current contact interactions.
We use the $C_{lq}^{(\pm)}$ combinations in the following because they map directly onto the down-type and up-type left-handed contact amplitudes.

With these definitions, the differential spin- and colour-averaged partonic cross section for $q_i \bar q_i \to \ell_k^+ \ell_k^-$ is
\begin{equation}
\frac{d\hat \sigma^{q_i \ell_k}}{d \hat t}
=
\frac{1}{48 \pi \hat{s}^2}
\left[
\hat{u}^2
\left(
|F_{q_L \ell_L}^{ik}(\hat{s})|^2
+
|F_{q_R \ell_R}^{ik}(\hat{s})|^2
\right)
+
\hat{t}^2
\left(
|F_{q_L \ell_R}^{ik}(\hat{s})|^2
+
|F_{q_R \ell_L}^{ik}(\hat{s})|^2
\right)
\right]~,
\label{eq:partonic_xsec}
\end{equation}
where $\hat{s} = m_{\ell \ell}^2$, $\hat{t}$, and $\hat{u}$ are the partonic Mandelstam variables.
Convolving Eq.~\eqref{eq:partonic_xsec} with PDFs and summing over quark flavours gives the leading-order SMEFT correction in each bin of $m_{\ell\ell}$ and $\cos\theta^*$.
The explicit expression is reported in App.~\ref{app:xsec}.

Since the amplitude is linear in the SMEFT coefficients, the cross section in each bin is a quadratic polynomial in the Wilson coefficients.
In the massless limit there is no interference between different fermion flavours or chiralities.
Therefore, in the basis of Eq.~\eqref{eq:contact_interactions}, the quadratic contribution contains no mixed products of different coefficients and we can write the cross section in each bin $I$ as
\begin{equation}
  \sigma^{I}_{\rm DY-SMEFT}({\bf C})
  =
  \sigma^{I}_{\rm DY-SM}
  \left[
    1
    +
    \sum_x a_x^{I} C_x
    +
    \sum_x b_x^{I} C_x^2
  \right]~,
  \label{eq:smeft_bin_xsec}
\end{equation}
where $a_x^{I}$ and $b_x^{I}$ encode the interference and quadratic contributions of the coefficient $C_x$.
The coefficients $C_x$ are taken to be real, as appropriate for the flavour-diagonal Hermitian operators considered here.

Keeping the quadratic terms in Eq.~\eqref{eq:smeft_bin_xsec} corresponds to computing $|\mathcal A_{\rm SM}+\mathcal A_{\rm dim6}|^2$, namely truncating the amplitude at dimension six and then squaring it.
This is not a complete SMEFT prediction at order $\Lambda^{-4}$, since dimension-eight interference terms are not included.
Nevertheless, for high-energy Drell-Yan processes this amplitude-level truncation often provides a better approximation to explicit UV completions than an observable-level truncation retaining only the linear interference term \cite{Allwicher:2024mzw,Brivio:2022pyi}.
We therefore use the linear-plus-quadratic prediction as our default.

\subsection{PDF dependence}
\label{sec:PDF}

The CMS covariance matrix includes several sources of systematic uncertainty in the background prediction, among which PDF uncertainties are sizeable in the high-mass bins. These PDF uncertainties, however, have already been propagated by CMS to the event-yield level for a fixed SM prediction and a fixed PDF-error prescription. They therefore enter only as bin-to-bin yield uncertainties, with the flavour and kinematic information of the underlying PDF variations already projected out. Such a covariance cannot be used to profile the flavour-dependent PDF degrees of freedom relevant for our analysis. We therefore introduce an independent PDF-nuisance parametrisation and use it to propagate PDF deformations consistently to the DY signal region.

Ideally, one would remove the PDF component from the CMS covariance before adding the explicit PDF nuisance parameters. However, since a breakdown of the different sources of covariance is not available, this is not possible. We therefore keep the CMS covariance unchanged, see also \cite{NNPDF:2021njg} for a similar approach. This makes the absolute constraints somewhat conservative, while it can mildly underestimate the relative impact of PDF profiling, as part of the PDF uncertainty is already present in the fixed-PDF likelihood as an effective yield covariance. For the HL-LHC projections this effect is further reduced by our rescaling of the Run-II systematic covariance by the factor \(r_{\rm syst}=0.2\), which is used in the baseline scenario.

For the PDF input to our computation of DY we adopt the `DIS+DY (no HM)' set introduced in Ref.~\cite{Greljo:2021kvv}.
This is an NNPDF-style SM Monte Carlo replica fit based on DIS and Drell-Yan data, excluding the high-mass neutral-current Drell-Yan measurements included in the corresponding DIS+DY baseline fit.
It therefore includes DIS structure-function data, fixed-target Drell-Yan measurements, Tevatron data, and LHC low-mass and on-shell electroweak-boson production.
The motivation for this choice is to use a PDF prior that is independent of the high-mass DY data entering the SMEFT fit.
Since contact interactions can induce energy-growing distortions in the same dilepton tails that constrain the large-$x$ quark PDFs, including those data in the PDF determination could partially absorb a possible SMEFT signal into the PDFs~\cite{Hammou:2023heg,Hammou:2024xuj}.
The `DIS+DY (no HM)' set avoids this overlap and provides a conservative reference for assessing the impact of PDF uncertainties on high-mass Drell-Yan SMEFT bounds.

The original Monte Carlo replica set is converted into a Hessian representation, following the standard MC2Hessian construction described in App.~\ref{app:pdf}.
This provides a compact nuisance-parameter description of the PDF uncertainty, in which the replica covariance is represented by a finite set of orthogonal Hessian directions that can be profiled in the likelihood.
For each parton $p$, the resulting PDF set consists of a central member, $f^{(0)}_p(x,Q)$, and $N_{\rm eig}$ Hessian directions, which we denote by the shifts
\begin{equation}
  \delta f^{(\alpha)}_p(x,Q) =
  f^{(\alpha)}_p(x,Q)-f^{(0)}_p(x,Q)~,
\end{equation}
where $\alpha=1,\ldots,N_{\rm eig}$ labels the Hessian direction.
A generic PDF configuration in the Hessian approximation is then written as
\begin{equation}
  f_p(x,Q;\boldsymbol{\lambda}) =
  f^{(0)}_p(x,Q) +
  \sum_{\alpha=1}^{N_{\rm eig}}
  \lambda_\alpha\,\delta f^{(\alpha)}_p(x,Q)~,
  \label{eq:pdf_hessian_description}
\end{equation}
where the nuisance parameters $\lambda_\alpha$ are assigned a Gaussian prior with zero mean and unit variance.

We propagate the PDF dependence to the binned DY cross section by recomputing the hadronic convolution for the central PDF member and for each Hessian direction.
Keeping only the linear dependence on the PDF nuisance parameters, the SM DY cross section in bin $I$ is parameterised as
\begin{equation}
  \sigma^{I}_{\rm DY-PDF}(\boldsymbol{\lambda})
  =
  \sigma_{\rm DY-SM}^{I\,(0)}
  \left(
    1
    +
    \sum_{\alpha=1}^{N_{\rm eig}} w^{I}_\alpha\,\lambda_\alpha
  \right)~,
  \label{eq:pdf_linearized_xsec}
\end{equation}
where $\sigma_{\rm DY}^{I\,(0)}$ is the prediction obtained with the central PDF member, and the numerical coefficients $w^{I}_\alpha$ are the relative shifts induced by each Hessian direction.
This linearised treatment is sufficient for the profiling performed in this work, since the fitted PDF nuisance parameters remain within the region described by the Hessian prior.

\subsection{Statistical analysis}
\label{sec:stat}

Since the expected deviations from the SM are small compared to the SM contribution, we neglect mixed SMEFT-PDF terms and evaluate the SMEFT templates with the central PDF member.
With this approximation, the PDF dependence enters only through the SM DY component, while the SMEFT and PDF corrections add linearly at the level of the relative DY prediction.
For each bin $I$ we define the ratio
\begin{equation}
    R_I({\bf C}, \boldsymbol{\lambda}) \equiv \frac{\sigma^{I}_{\rm DY-SMEFT+PDF}}{\sigma^{I\,(0)}_{\rm DY-SM}} \approx 1 + \sum_x \left( a_x^{I} \, C_x + b_x^{I} \, C_x^2 \right) + \sum_\alpha w_\alpha^{I} \, \lambda_\alpha ~.
    \label{eq:ratio_template}
\end{equation}
The expected central event yield in the CMS Run-II dataset is then obtained by rescaling only the DY component of the CMS background prediction of Eq.~\eqref{eq:total_sm},
\begin{equation}
    \mu_{I}({\bf C}, \boldsymbol{\lambda}) = R_{I}({\bf C}, \boldsymbol{\lambda}) \, N^{\rm DY, CMS}_{I} + N^{\text{non-DY, CMS}}_{I}\, .
\end{equation}
This retains the higher-order QCD and electroweak corrections included in the CMS prediction, under the assumption that these corrections factorise to good approximation from the SMEFT and PDF deformations considered here.
For the HL-LHC projections we use the analogous expression
\begin{equation}
    \mu_{I}^{\rm HL}({\bf C}, \boldsymbol{\lambda}) = K_{I}^{\rm HL} \left( R_{I}^{\rm HL}({\bf C}, \boldsymbol{\lambda})  \, N^{\rm DY, CMS}_{I} + N^{\text{non-DY, CMS}}_{I} \right)\, ,
\end{equation}
where $R_{I}^{\rm HL}({\bf C},\boldsymbol{\lambda})$ is evaluated for $pp$ collisions at $\sqrt{s}=14~{\rm TeV}$, instead of $13~{\rm TeV}$.

Correlated experimental systematic uncertainties are implemented through log-normal nuisance parameters.
Let $\mu_{I}^0 \equiv N_{I}^{\rm DY}+N_{I}^{\rm non-DY}$ denote the central SM expectation around which the experimental covariance is defined, and let $V_{IJ}$ be the corresponding covariance matrix for the background prediction.
We map this yield covariance to a latent Gaussian covariance through
\begin{equation}
  \Sigma_{IJ} = \log\left(1+\frac{V_{IJ}}{\mu_{I}^0\mu_{J}^0}\right) \, .
  \label{eq:lognormal_cov}
\end{equation}
For standard-normal experimental nuisance parameters $\xi_a$, the full
expected yield entering the likelihood is
\begin{equation}
  \mu_{I}({\bf C}, \boldsymbol{\lambda}, \boldsymbol{\xi})
  =
  \mu_{I}^{\rm cent}({\bf C}, \boldsymbol{\lambda})
  +
  \mu_{I}^0
  \left[
    \exp\!\left((L \, \xi)_{I}\right)-1
  \right] \, .
  \label{eq:mu_lognormal}
\end{equation}
The $n_{\rm Bins} \times n_{\text{exp-nuis}}$ matrix $L$ is defined by the decomposition $\Sigma=L L^T$.
In the CMS aggregated systematics, the number of nuisance parameters $n_{\text{exp-nuis}}$ describing systematic uncertainties is equal to the number of bins $n_{\rm Bins}$.
This construction reproduces the correlated covariance $V$ to first order around the SM prediction, while keeping the nuisance-induced yield shifts multiplicative in the underlying log-normal variables.

The full test statistic is
\begin{equation}
  \chi^2({\bf C}, \boldsymbol{\lambda}, \boldsymbol{\xi})
  =
  2\sum_{I}
  \left[
    \mu_{I} ({\bf C}, \boldsymbol{\lambda}, \boldsymbol{\xi})
    - n_{I}
    + n_{I} \log\left(\frac{n_{I}}{\mu_{I} ({\bf C}, \boldsymbol{\lambda}, \boldsymbol{\xi})}\right)
  \right]
  +
  \sum_\alpha \lambda_\alpha^2
  +
  \sum_a \xi_a^2 ~.
  \label{eq:full_chi2}
\end{equation}
The logarithmic term is set to zero for empty observed bins, corresponding to the $n\log n \to 0$ limit.
For each point in Wilson-coefficient space, the PDF and experimental nuisance parameters are profiled:
\begin{equation}
  \chi^2_{\rm prof}({\bf C})
  =
  \min_{\boldsymbol{\lambda}, \boldsymbol{\xi}}
  \chi^2({\bf C}, \boldsymbol{\lambda}, \boldsymbol{\xi}) ~ .
  \label{eq:profchi2}
\end{equation}
Confidence intervals are obtained from $\Delta\chi^2({\bf C})=\chi^2_{\rm prof}({\bf C})-\chi^2_{\rm prof}({\bf \hat{C}})$, where ${\bf \hat{C}}$ denotes the best-fit point.
We use Wilks' approximation and take $\Delta\chi^2=3.84$ for one-parameter 95\% CL intervals and $\Delta\chi^2=5.99$ for two-parameter 95\% CL contours.

The HL-LHC projection is analysed with the same likelihood and nuisance parameter treatment. The observed counts, central expectations, and experimental covariance are replaced by the projected Asimov inputs described in Sec.~\ref{subsec:hl_data}, while the SMEFT ratio templates and PDF nuisance directions enter as in Eq.~\eqref{eq:ratio_template}.

The public results repository
\textcolor{black}{\faGithub}\ 
\href{https://github.com/manuel-morales-a/flavourful-drell-yan-smeft-results}
{\texttt{flavourful-drell-yan-smeft-results}} contains a Mathematica notebook with an analytic approximation for the profiled $\chi^2_{\rm prof}({\bf C})$ as a quartic polynomial, for the three scenarios of LFU, electron coefficients only, and muon coefficients only, for both current CMS data and HL-LHC projections.

\section{Results}
\label{sec:results}

In this section we present the SMEFT fits to the current CMS high-mass Drell-Yan data \cite{CMS:2021ctt}, together with the corresponding HL-LHC projections.
Our main goal is to quantify how PDF uncertainties affect high-energy dilepton constraints in scenarios with different quark-flavour structures.
Since the PDF dependence is common to the dielectron and dimuon channels, while lepton-flavour non-universal effects can be disentangled by comparing the two final states, we focus in the main text on lepton-flavour-universal (LFU) scenarios.
In these fits the SMEFT coefficients entering the electron and muon channels are assumed to be equal.
The fully flavour-general results, where this assumption is lifted, are collected in App.~\ref{app:fl_general}.

For each scenario we compare constraints obtained with the PDF nuisance parameters fixed to zero to those obtained after profiling them in the likelihood.
In both cases the experimental nuisance parameters are treated in the same way; the difference therefore isolates the impact of PDF profiling.
To quantify this effect we define the \emph{broadening} factor as
\begin{equation}
C\text{-Broadening}_{\rm PDF} \equiv
\frac{\delta C^{\rm PDF}}
     {\delta C^{\rm No\text{-}PDF}}
-1 \, ,
\label{eq:pdf_broadening}
\end{equation}
where $\delta C^{\rm PDF}$ and $\delta C^{\rm No\text{-}PDF}$ denote the 95\% CL interval widths obtained with profiled and fixed PDF nuisance parameters, respectively.
A positive broadening therefore indicates a weakening of the SMEFT constraint due to profiling over PDF uncertainties.

\subsection{Universal $W$ and $Y$ parameters}
\label{subsec:wy_basis}

To compare with available results in the literature, we start with a fit of the universal $W$ and $Y$ parameters \cite{Barbieri:2004qk}. In the SMEFT, these can be defined as the coefficients of the dimension-six operators
\begin{equation}
    \mathcal L_{\rm SMEFT}^{W,Y} = - \frac{W}{2 m_W^2} (D^\mu W_{\mu \nu}^a)^2 - \frac{Y}{2 m_W^2} (\partial^\mu B_{\mu \nu})^2~,
\end{equation}
where $W^a_{\mu\nu}$ and $B_{\mu\nu}$ are the $SU(2)_L$ and $U(1)_Y$ field strengths, respectively, and $D^\mu$ is the covariant derivative. By construction, this scenario is flavour universal in both the quark and lepton sectors.
Using the SM equations of motion for the gauge fields, the $W$ and $Y$ scenario can be written in the Warsaw basis. The resulting semileptonic operators, relevant to our analysis, are generated in flavour-universal combinations
\begin{equation}
\begin{aligned}
\mathcal L_{\rm SMEFT}^{W,Y}
\supset
&-
\frac{g^2W}{4m_W^2}\,
\mathcal O_{lq}^{(3)}
-
\frac{g_Y^2Y}{m_W^2}
\bigg(
Y_lY_d\,\mathcal O_{ld}
+Y_lY_u\,\mathcal O_{lu}
\\
&+Y_lY_q\,\mathcal O_{lq}^{(1)}
+Y_eY_d\,\mathcal O_{ed}
+Y_eY_u\,\mathcal O_{eu}
+Y_eY_q\,\mathcal O_{qe}
\bigg)~.
\end{aligned}
\label{eq:wy_lagrangian}
\end{equation}
where $Y_l=-1/2$, $Y_e=-1$, $Y_q=1/6$, $Y_u=2/3$, and $Y_d=-1/3$ are the SM fermion hypercharges, and flavour indices are suppressed.

\begin{figure}[t]
    \centering
    \includegraphics[width=0.9\linewidth]
    {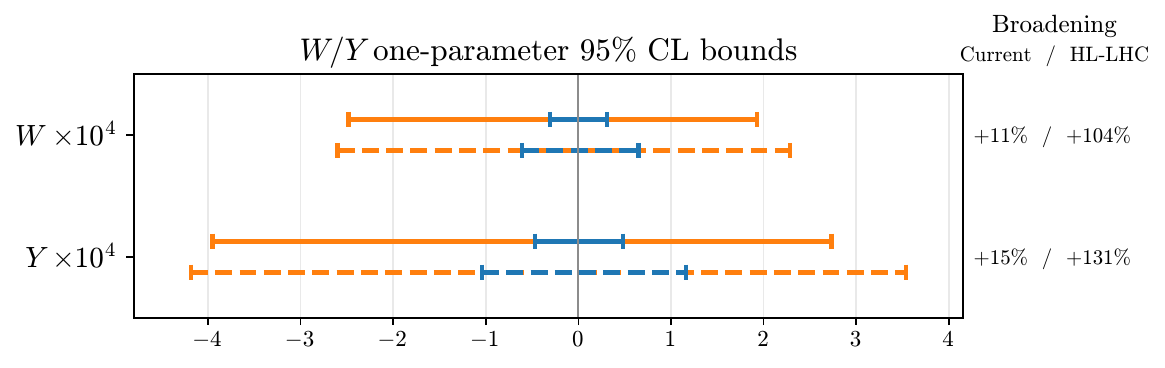}\\
    \hspace{-1.5 cm}\includegraphics[width=0.7\linewidth]{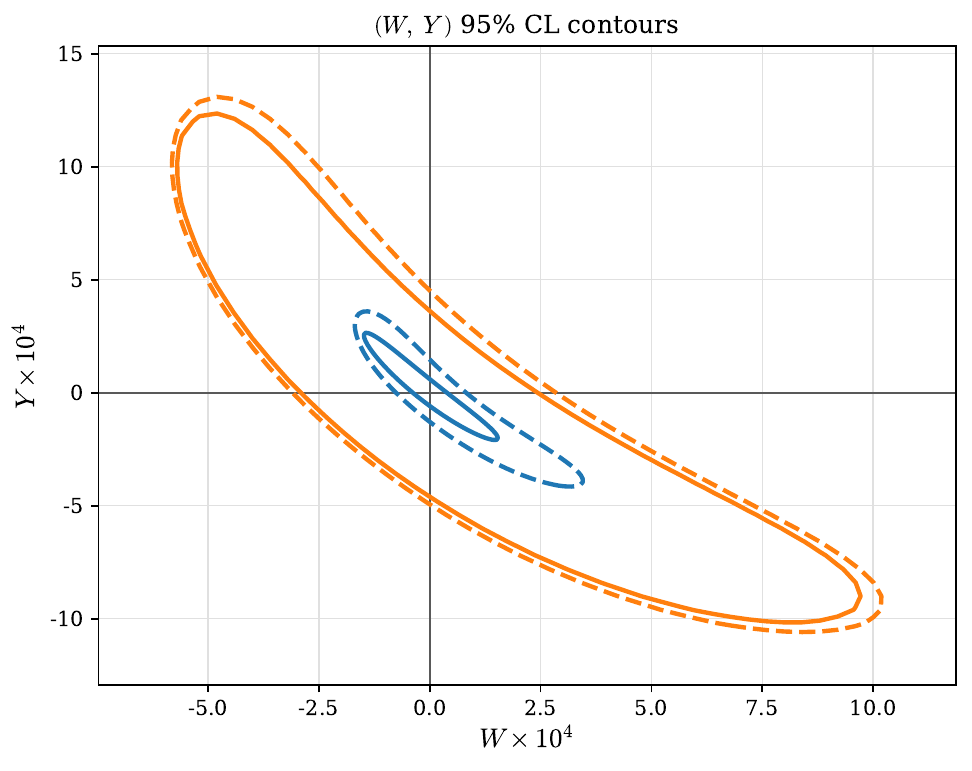}\\
    \includegraphics[width=0.6\linewidth]{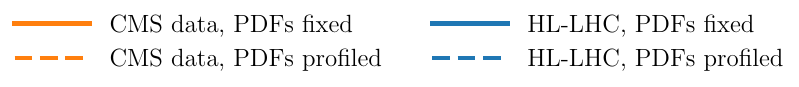}
    \caption{One-parameter intervals (top) and two-parameter contours (bottom) for the $W$ and $Y$ parameters at 95\% CL, using current CMS data (orange) and HL-LHC projections (blue). Solid intervals and contours correspond to fixed-PDF fits, while dashed ones include PDF profiling.}
    \label{fig:wy_1d_current_data}
\end{figure}

The current constraints from CMS data and the projected HL-LHC sensitivities in the $(W,Y)$ scenario are shown in Figure~\ref{fig:wy_1d_current_data}. The top panel describes single-parameter fits, while the bottom plot shows the simultaneous fit of $W$ and $Y$. Results from current CMS data are shown in orange, while HL-LHC projections are displayed in blue. Solid lines correspond to fixed-PDF fits, while dashed ones include profiling over PDF nuisance parameters.

For the one-parameter fits, the broadening induced by PDF profiling is $11\%$ for $W$ and $15\%$ for $Y$ with current data. At the HL-LHC the same broadening increases to $104\%$ and $131\%$, respectively.

Our results for the present constraints and HL-LHC projections for the $W$ and $Y$ parameters are qualitatively compatible with those reported in Refs.~\cite{Torre:2020aiz,Greljo:2021kvv}.
Exact numerical agreement is not expected, since we use a different CMS dataset and a different binning strategy, including angular information but with coarser invariant-mass bins (the HL-LHC projection with merged angular bins is shown in Figure~\ref{fig:2D_angular_information}).
The largest difference appears in the projected HL-LHC sensitivity to $W$ in the fixed-PDF case, which is stronger in Ref.~\cite{Greljo:2021kvv} because their HL-LHC projection also includes charged-current Drell-Yan data.
The projected sensitivity to $Y$, as well as the size of the PDF broadening, is instead closer to our result.

In particular, the comparison with \cite{Greljo:2021kvv} provides a useful validation of our simplified PDF treatment.
It shows that profiling a Hessian PDF nuisance basis captures the dominant PDF variability relevant for high-mass Drell-Yan SMEFT fits, at least at the level needed for the phenomenological comparisons performed here.

\subsection{Lepton flavour universal scenario}
\label{subsec:lfu_scenario}

\begin{figure}[t]
    \centering
    \includegraphics[width=0.97\linewidth]{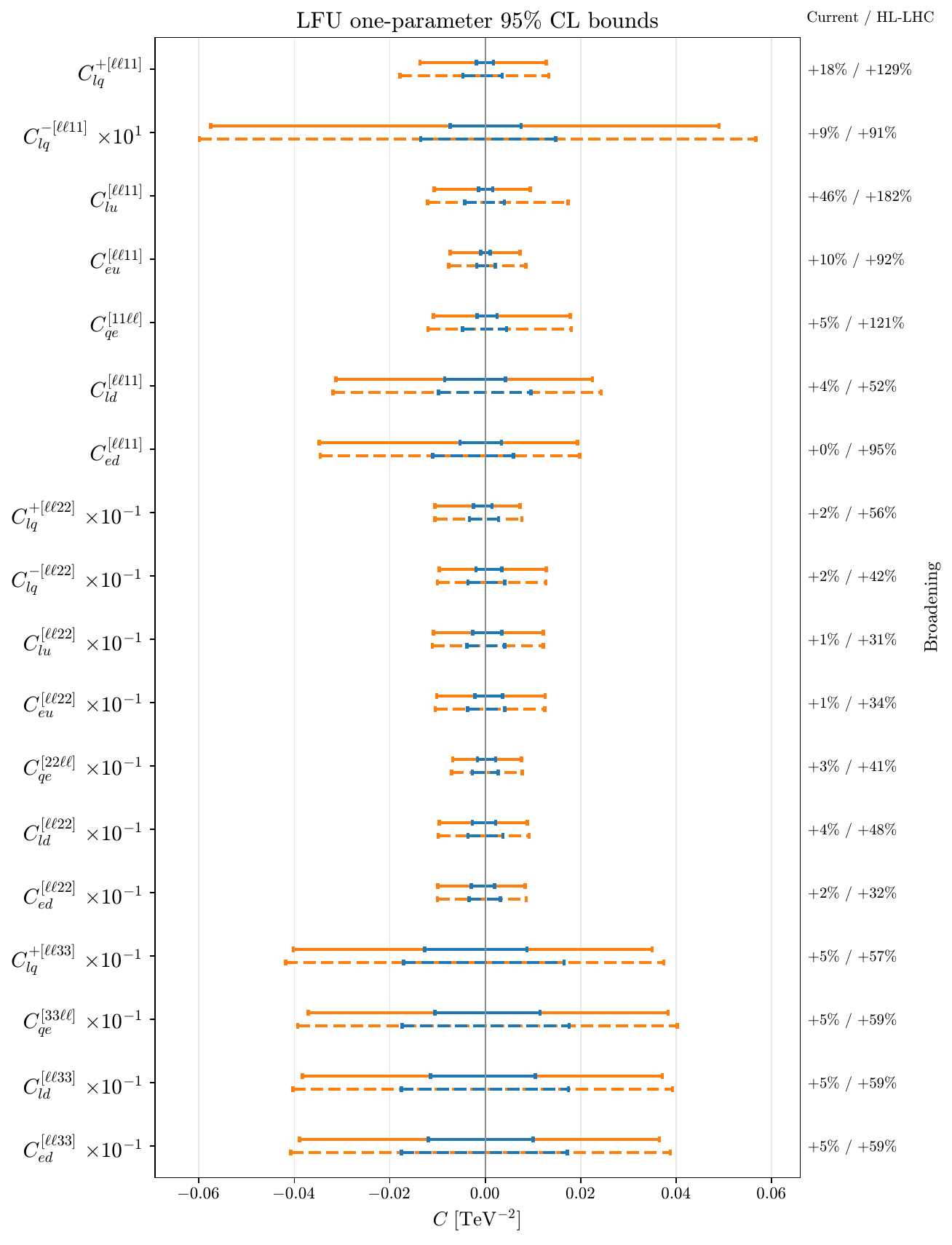} \\
    \includegraphics[width=0.6\linewidth]{figures/2D_LFU_legend.pdf}
    \caption{One-parameter LFU bounds from current data and HL-LHC sensitivity projections. The percentage labels indicate the relative change in the interval width induced by PDF profiling.}
    \label{fig:lfu_1D_results}
\end{figure}

We now turn to the LFU scenario, with the goal of studying how the impact of PDF profiling depends on the quark flavour structure of the SMEFT operators.
In this setup the Wilson coefficients in the electron and muon channels are identified, while the quark-flavour dependence is kept fully resolved.

\begin{figure}[t]
    \centering
    {\small LFU two-parameter 95\% CL contours}\\
    \includegraphics[width=0.43\linewidth]{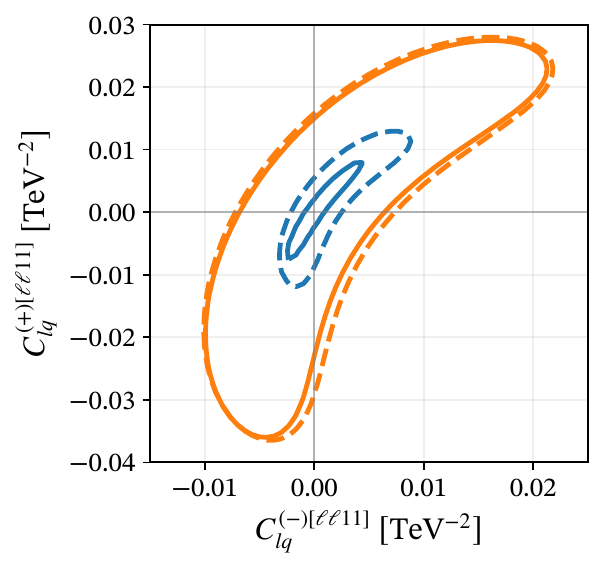}
    \includegraphics[width=0.43\linewidth]{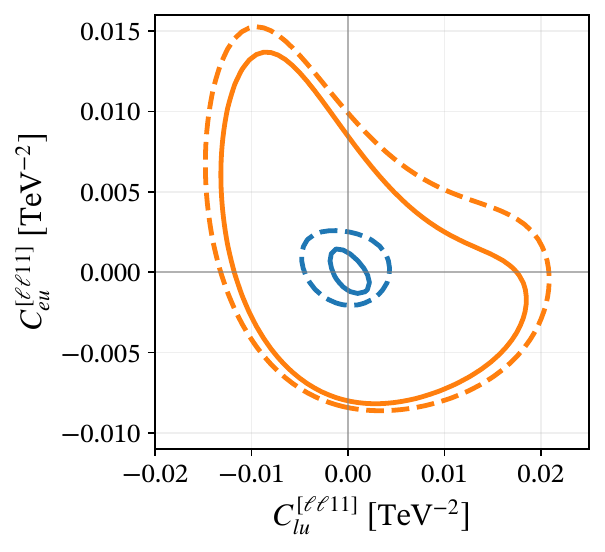} \\
    \includegraphics[width=0.43\linewidth]{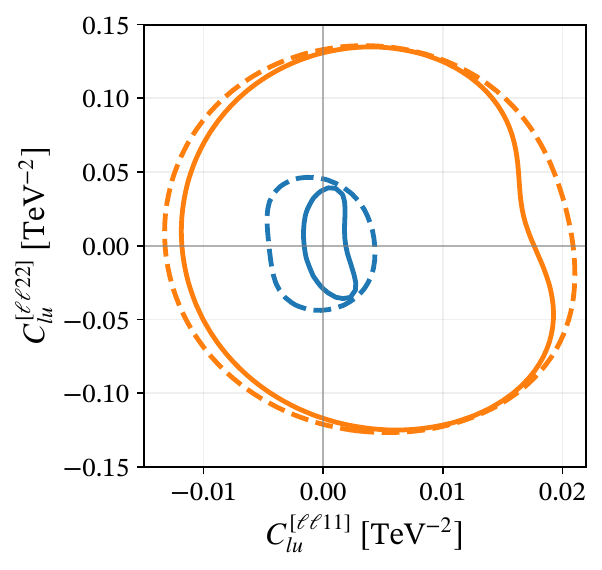}
    \includegraphics[width=0.43\linewidth]{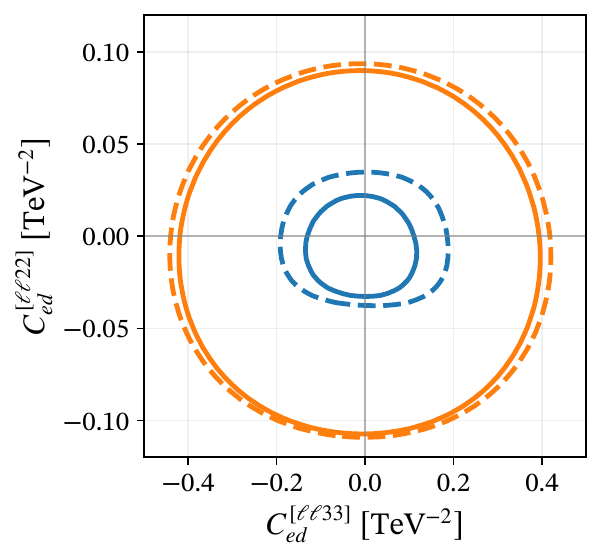} \\
    \includegraphics[width=0.6\linewidth]{figures/2D_LFU_legend.pdf}
    \caption{Two-coefficient 95\% CL contours in the LFU scenario, using current data (orange) and HL-LHC sensitivity projections (blue). Solid lines correspond to fixed PDFs while dashed ones implement PDF profiling.}
    \label{fig:lfu_2D_results}
\end{figure}

The current one-parameter 95\% CL bounds from CMS data (orange) and the future HL-LHC sensitivities (blue) are shown in Figure~\ref{fig:lfu_1D_results}.
In Figure~\ref{fig:lfu_2D_results} we show two-parameter 95\% CL contours for a representative selection of coefficient pairs.
In all plots, solid and dashed lines correspond to fits in which the PDF nuisance parameters are fixed to zero or profiled, respectively.
A clear flavour dependence of the PDF-profiling effect is visible, both with current data and in the HL-LHC projection.

The largest broadening occurs for operators involving first-generation quarks.
For these directions the effect is typically of order $10\%$ with current data and can reach order $100\%$ at the HL-LHC.
Operators involving second- and third-generation quarks show smaller broadenings, typically at the level of a few percent with current data and around $50\%$ in the HL-LHC projection.
There is, however, some operator-dependent variability.
For example, the current-data broadening for $C_{lu}^{[\ell\ell 1 1]}$ reaches $46\%$, due to the interplay between the interference and quadratic SMEFT terms, which distorts the $\chi^2$ profile at positive values of the coefficient.
Conversely, the broadening of $C_{ed}^{[\ell\ell 1 1]}$ with current data is very small, likely because of an accidental cancellation in the relevant bins.

The two-parameter contours in Figure~\ref{fig:lfu_2D_results} show sizeable correlations for first-generation operators, where both the interference and quadratic SMEFT contributions are important.
For operators involving heavier quarks, the sensitivity is driven more strongly by the quadratic terms, and the correlations tend to be reduced.

The hierarchy of broadening between first-generation and heavier-quark operators follows the flavour structure of the partonic channels that dominate high-mass Drell-Yan.
The strongest effect is found for first-generation light-quark operators, whose contributions are weighted by the high-$x$ $u\bar u$ and $d\bar d$ luminosities.
These are precisely the directions where PDF uncertainties are most relevant.
When the PDF nuisance parameters are profiled, part of the spectral distortion induced by these SMEFT coefficients can therefore be absorbed by shifts in the PDFs, leading to broader intervals.
This degeneracy is weaker for operators involving second- and third-generation quarks, whose parton luminosities are not valence-enhanced in the same kinematic region.

\begin{figure}[t]
    \centering
    \includegraphics[width=0.95\linewidth]{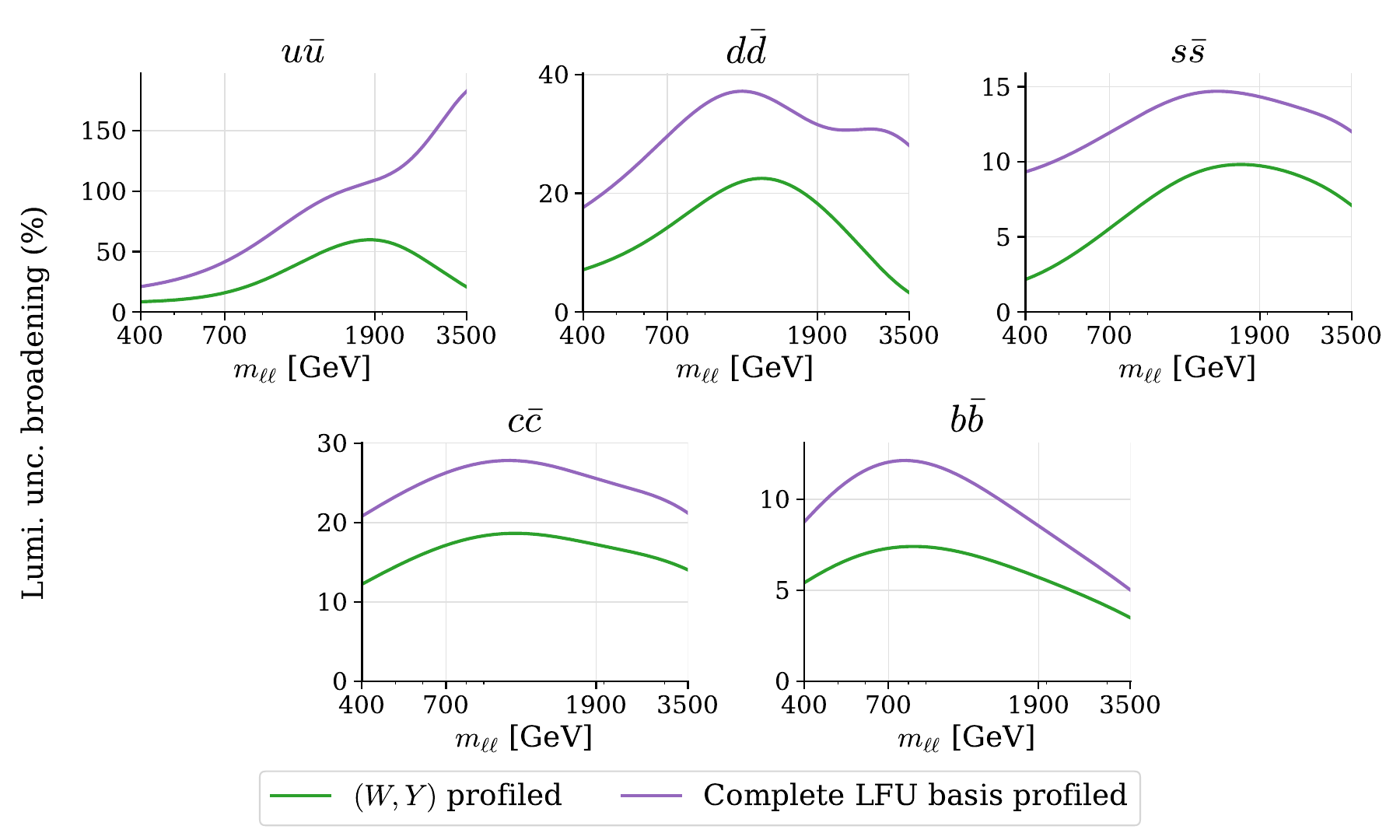}
    \caption{PDF broadening for different parton luminosities at the HL-LHC with respect to the SM PDF fit.}
    \label{fig:hllhc_lfu_lumi_broad}
\end{figure}

A complementary way to quantify the SMEFT-PDF interplay is to study the broadening of the PDF posterior uncertainties when SMEFT coefficients are profiled.
At the level of parton luminosities, we define this broadening from the ratio of luminosity uncertainties,
\begin{equation}
  \text{PDF-Broadening}_{\rm SMEFT} \equiv
  \frac{
    \sigma\!\left[\mathcal{L}_{q\bar q}^{\rm SMEFT-prof}(m_{\ell\ell})\right]
  }{
    \sigma\!\left[\mathcal{L}_{q\bar q}^{\rm SM}(m_{\ell\ell})\right]
  }-1 ~.
\end{equation}
Figure~\ref{fig:hllhc_lfu_lumi_broad} shows this quantity after profiling either all LFU SMEFT coefficients or only the flavour-universal $(W,Y)$ directions.
The general LFU scenario has a substantially larger impact on the posterior PDF uncertainties.
For some luminosities, the broadening is about twice as large as in the flavour-universal $(W,Y)$ scenario.
The $u\bar u$ luminosity is the most affected in both cases; in the LFU scenario its uncertainty increases by up to a factor of three, corresponding to a $200\%$ broadening, in the large invariant-mass region.
This broadening occurs in a kinematic region where the PDFs are still meaningfully constrained, as shown in Figure~\ref{fig:hllhc_lfu_wy_lumi_bands} in App.~\ref{app:pdf_prior}.
The large enhancement of the $u\bar u$ uncertainty is therefore not simply an artefact of an extrapolation region where the baseline PDF uncertainty is already very large.
It should instead be interpreted as a genuine loss of PDF constraining power from high-mass Drell-Yan data once SMEFT deformations are allowed.

Finally, we quantify the impact of the assumptions on experimental systematic uncertainties.
Comparing the SMEFT HL-LHC sensitivities obtained under the optimistic assumption $r_{\rm syst}=0.2$, which is our baseline result, with those using current systematic uncertainties, $r_{\rm syst}=1$, we find that the PDF-profiled 95\% CL interval widths increase by roughly $10$--$20\%$ depending on the specific coefficient.
Moreover, we have checked that the relative broadening induced by PDF profiling is smaller in the scenario with larger systematics. This reflects the fact that the impact of PDF uncertainties is diluted when other sources of uncertainty become larger.

\subsubsection*{Specific case study}

\begin{figure}[t]
  \centering
  \includegraphics[width=0.4\textwidth]{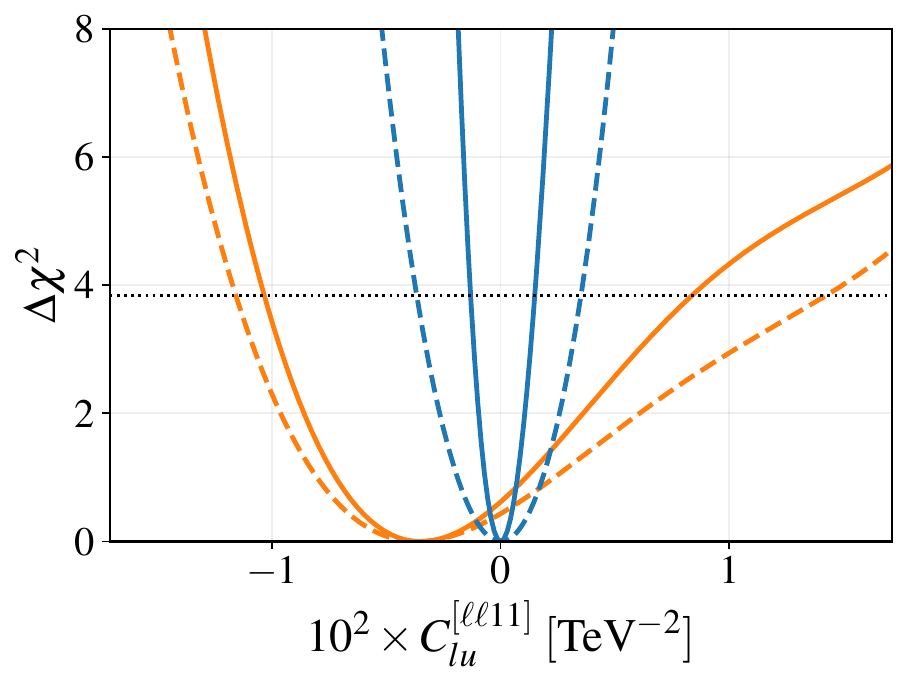}
  \quad
  \includegraphics[width=0.4\textwidth]{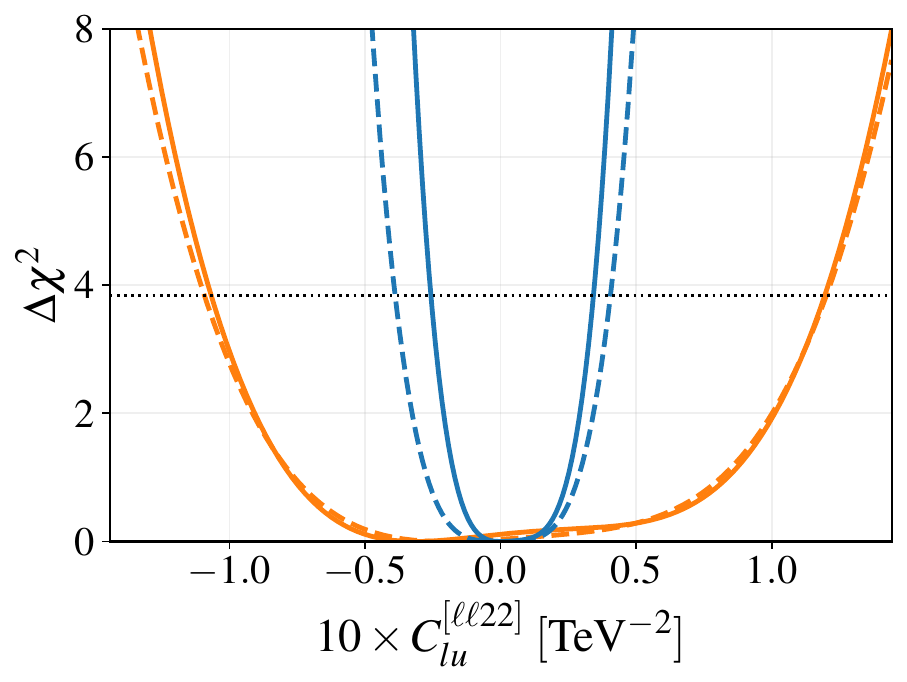} \\
  \includegraphics[width=0.6\linewidth]{figures/2D_LFU_legend.pdf}\\
  \includegraphics[width=0.95\linewidth]{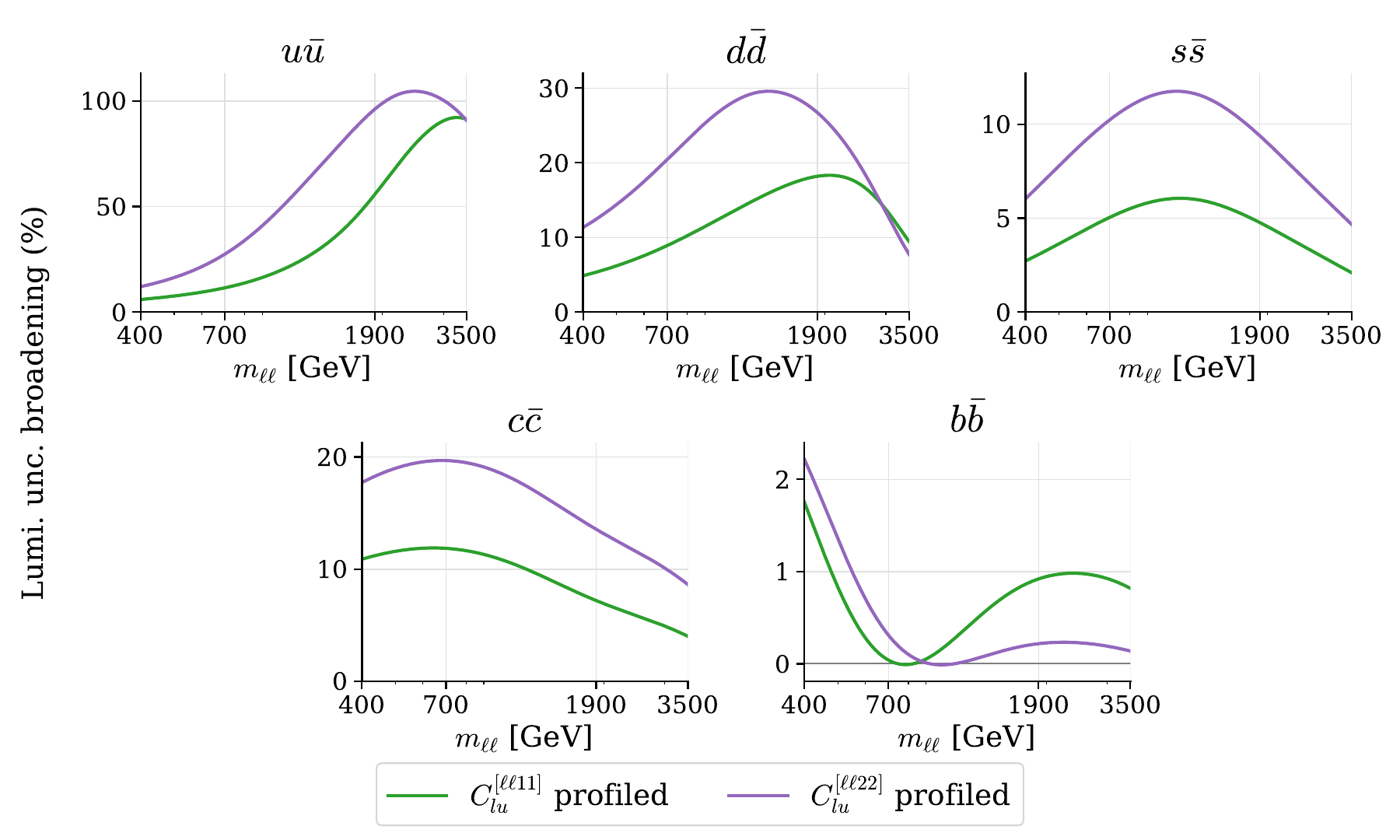}
  \caption{{\bf Top:} Quadratic LFU $\Delta\chi^2$ profiles for current data and HL-LHC, with and without PDF profiling, for $C_{lu}^{[\ell\ell 11]}$ (left) and $C_{lu}^{[\ell\ell 22]}$ (right). The horizontal line shows the 95\% $\Delta \chi^2$ threshold.
  {\bf Bottom:} PDF luminosities broadening for different parton channels at the HL-LHC with respect to the SM PDF fit.}
  \label{fig:clu-chi2-pdf-profiling-hllhc}
\end{figure}

Let us now explore in more detail the source of the different behaviour we observe between a first-generation and a second-generation operator. For our example, we focus specifically on the $C_{lu}^{[\ell\ell 1 1]}$ and $C_{lu}^{[\ell\ell 2 2]}$ coefficients, as they show the largest difference in the broadening factors among generations.
In our one-parameter fits, Figure~\ref{fig:lfu_1D_results}, we obtain a broadening of 46\% with current data and 182\% at HL-LHC for the first-generation operator, whereas for the second-generation operator the broadening is only 1\% with CMS data and 31\% at HL-LHC. This behaviour can also be observed from the bottom-left panel of Figure~\ref{fig:lfu_2D_results}.

In Figure~\ref{fig:clu-chi2-pdf-profiling-hllhc} we plot the $\Delta \chi^2$ curves for CMS data (orange) and HL-LHC (blue) fits of the two coefficients, for both cases of fixed (solid) and profiled PDFs (dashed).
What we see from these plots is that in the case of positive values of $C_{lu}^{[\ell\ell 11]}$ (left panel) there is a strong interplay between the interference and the quadratic EFT contributions, which distorts the $\Delta \chi^2$ particularly in the fixed PDF scenarios. It is likely that this interplay gets strongly affected by profiling over PDF parameters, enhancing the broadening.
On the other hand, in the case of $C_{lu}^{[\ell\ell 2 2]}$ (right panel) the quadratic EFT term dominates over the interference, so that this interplay cannot take place and the broadening is similar to the one we obtain for the other second-generation coefficients.

Studying the effect that these two coefficients have in the $[1.9,3.5]$~TeV bins, setting their size at the corresponding $2\sigma$ bound, we see that $C_{lu}^{[\ell\ell 11]}$ induces an asymmetry between the $\cos \theta^* < 0$ and the $\cos \theta^* \geq 0$ bins of a factor $\sim 6.5$, while $C_{lu}^{[\ell\ell 2 2]}$ induces a smaller asymmetry of a factor $\sim 2$.
The PDF uncertainties follow a similar pattern, being about 13\% for $\cos \theta^* < 0$ and 3\% in $\cos \theta^* \geq 0$, i.e.\ a factor of $\sim 4$ between the two. This provides an additional explanation for why PDF uncertainties can better accommodate an effect from $C_{lu}^{[\ell\ell 11]}$ than from $C_{lu}^{[\ell\ell 2 2]}$.

The broadenings of the different parton luminosities in this LFU scenario are shown in the bottom panels of Figure~\ref{fig:clu-chi2-pdf-profiling-hllhc}. They display the relative increase of the PDF posterior uncertainties after profiling over the EFT coefficient, with respect to the SM PDF fit. The broadening is concentrated in the high-mass region and is the largest in the quark-antiquark luminosities that dominate the Drell-Yan tail, the valence sector. This provides the PDF-level counterpart of the broadening observed in the one-dimensional EFT bounds: profiling over an EFT deformation opens directions in PDF space that can partially accommodate changes in the high-mass dilepton spectrum, and the effect is most felt by the dominant luminosity channels.

\subsection{Impact of angular information}
\label{subsec:ang_information}

\begin{figure}[t]
    \centering
    \includegraphics[width=0.87\linewidth]{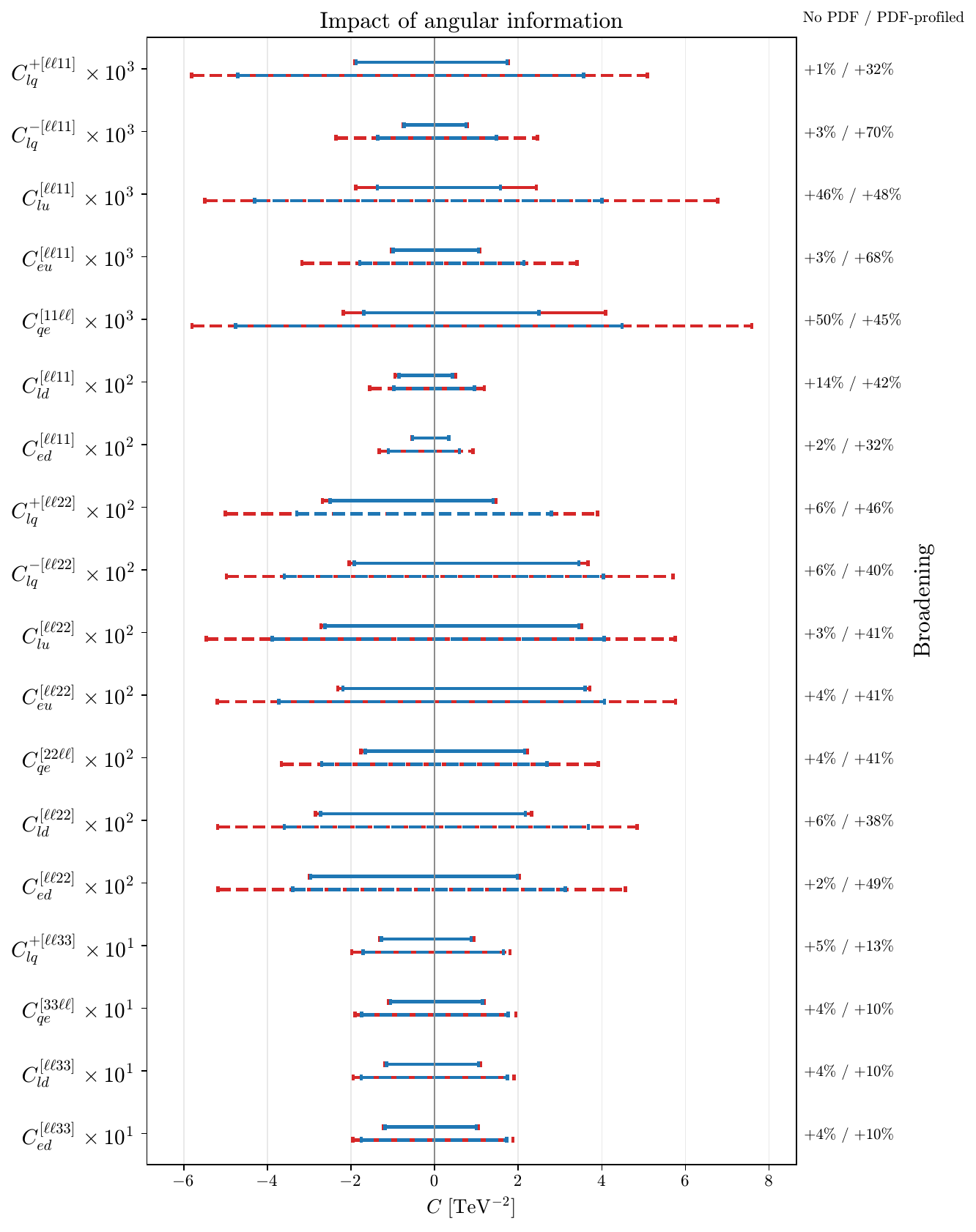}\\
    \includegraphics[width=0.8\linewidth]{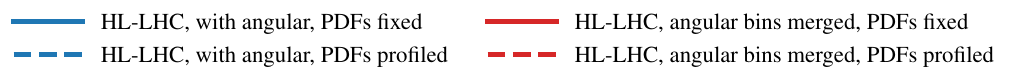}
    \caption{95\% CL sensitivity projections at the HL-LHC with angular information (blue) and integrating over the angle (red).}
    \label{fig:ang_info_hllhc}
\end{figure}

Including angular information in EFT studies of high-energy tails is useful to disentangle the effects of different operators. This is also true for dilepton tails, where operators with different chirality combinations induce different dependence on the partonic polar angle $\theta$, see Eq.~\eqref{eq:partonic_xsec}.
In this section we study a different aspect of angular information, specifically how much it contributes in disentangling EFT contributions from PDF effects. In fact, PDFs are chirality-independent and their convolution with the partonic cross sections does not induce an additional angular dependence. It is therefore reasonable to expect that angular information can help in reducing the broadening in EFT constraints when PDF nuisance parameters are profiled.

\begin{figure}[t]
  \centering
  \includegraphics[width=0.45\textwidth]{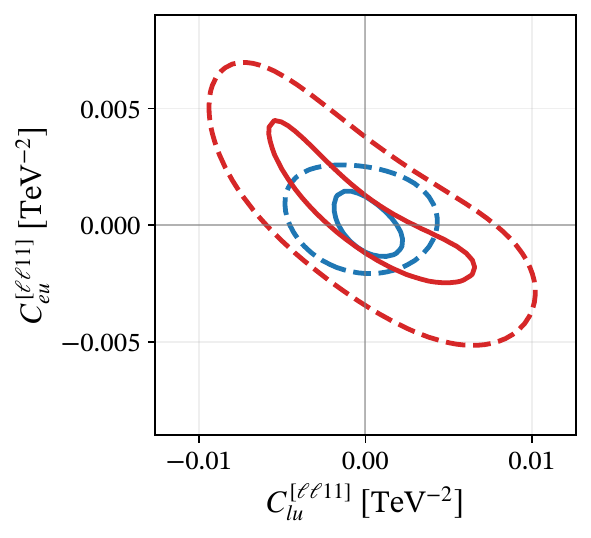}
  \hfill
  \includegraphics[width=0.45\textwidth]{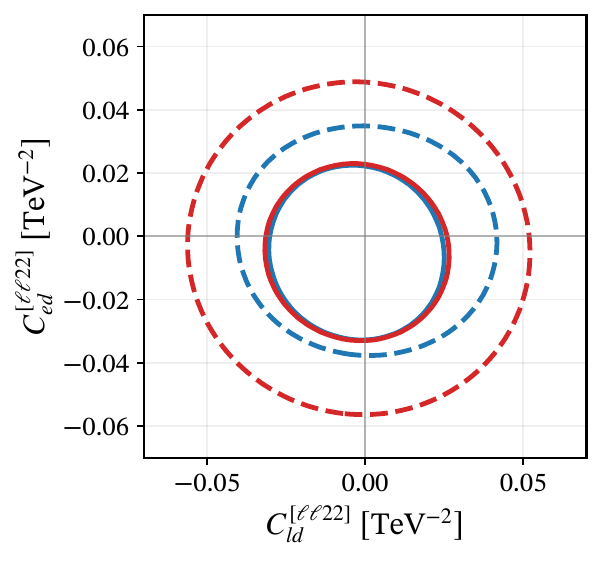} \\
  \includegraphics[width=0.40\textwidth]{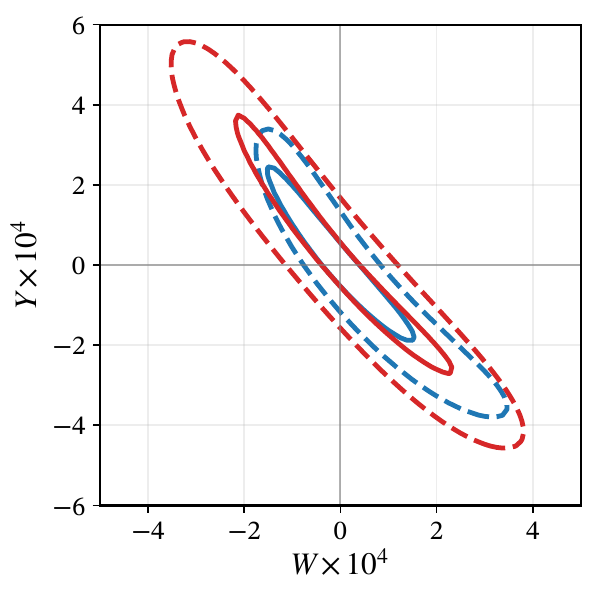}
  \\
  \includegraphics[width=0.8\linewidth]{figures/2D_LFU_angular_information_legend.pdf}
  \caption{Two-coefficient 95\% CL contours for HL-LHC differentiating (blue) or merging (red) angular bins. Solid lines correspond to fixed PDFs while dashed ones implement PDF profiling.}
  \label{fig:2D_angular_information}
\end{figure}

In Figure~\ref{fig:ang_info_hllhc} we show the one-parameter 95\% CL expected sensitivity interval at HL-LHC for EFT coefficients, obtained with the two $\cos \theta^*$ bins (blue) or merging them (red), both in the case of fixed PDFs (solid) and profiling over PDF nuisance parameters (dashed).
In the right column we report the broadening of the sensitivity interval induced by integrating over the angular variable, compared to when it is included in the fit, in both PDF-fixed and PDF-profiled cases, 
\begin{equation}
    C\text{-Broadening}_{\, \theta^*} \equiv
    \frac{\delta C_{{\rm No-}\theta^*}}{\delta C_{{\rm With-}\theta^*} } - 1~.
\end{equation}
We observe how the broadening due to the merging of the two angular bins is quite modest when PDFs are fixed. This can be expected since such a one-parameter fit relies mostly on the invariant mass dependence measured in the cross section.

The impact of angular information is also clearly evident in two-parameter fits.
In the top panels of Figure~\ref{fig:2D_angular_information} we show two examples of pairs of operators with same-flavour quark but different chiral structure. 
We see how the flat direction that appears when angular bins are merged in the first-generation fit (left panel), in which the interference dominates, is lifted once angular information is included. In the case of second-generation quarks (right panel) the quadratic terms dominate so no flat direction appears and in the EFT-only fit (solid lines) merging the angular bins or keeping them separate gives practically the same result. However, once PDFs are profiled, the fit with angular bins provides much better sensitivity.
In the bottom panel we show the impact of angular information in the $(W, Y)$ fit, finding that it substantially improves the constraints, both with fixed and profiled PDFs.

This observation provides further motivation for including an even finer angular binning in future experimental analyses of dilepton tails.

\section{Conclusions}
\label{sec:conclusions}

The program of pursuing flavour physics from high-energy probes is a relatively new addition to the more traditional one focused on low-energy processes. It has been shown that this program can be very powerful and can provide complementary information on the flavour structure of new physics. High-energy probes do not suffer from the challenges in computing matrix elements in non-perturbative low-energy QCD, but require care in dealing with other sources of theoretical uncertainties. Among those, PDFs represent the largest uncertainties for Drell-Yan studies at hadron colliders and in this paper we quantified their impact and proposed a simplified way to properly take them into account.

Specifically, we studied the role of PDF uncertainties in high-energy Drell-Yan constraints on SMEFT operators with different quark flavours. We have compared SMEFT sensitivities in fixed-PDF and PDF-profiled settings. These comparisons make it possible to isolate the degradation of the SMEFT sensitivity that is induced by PDF uncertainties. 

In the flavour-universal $(W, Y)$ scenario, our results show that profiling the Hessian PDF nuisance directions accounts for the dominant PDF effects relevant to the high-mass Drell-Yan fits considered here with respect to full PDF fits. Within the accuracy needed for the analyses presented here, our strategy provides a practical description of the PDF uncertainty entering the likelihood. Our findings are consistent with previous studies of universal SMEFT effects in high-mass Drell-Yan, where PDF uncertainties are found to have a sizeable but moderate effect in current data, and an increasingly important role at the HL-LHC. The same behaviour is visible in our projection: compared with current data, the profiled-PDF intervals are broadened more substantially relative to the fixed-PDF ones. This reflects the fact that, as more statistics are collected and the projected experimental sensitivity improves, the remaining freedom in the high-$x$ PDFs becomes a more important component of the likelihood and the SMEFT sensitivity.

We found that a rich pattern appears in the LFU scenario with arbitrary quark flavour. In this case, the impact of PDF profiling is visibly quark-flavour dependent, both for present data and in the HL-LHC projection. The largest broadenings are generally associated with operators involving first-generation quarks, while operators involving second- and third-generation quarks tend to show smaller effects, although with some operator-to-operator variation. This hierarchy is one of the main results of the analysis, since it shows that the relevance of PDF profiling is not uniform across the SMEFT flavour space.

The origin of this behaviour can be traced to the partonic channels that dominate high-mass Drell-Yan. First-generation light-quark operators receive enhanced contributions from the high-$x$ $u\bar u$ and $d\bar d$ luminosities. These are also the directions in which PDF uncertainties are most relevant for the high-mass spectrum, where valence channels dominate. When the PDF nuisance parameters are profiled, part of the spectral deformation produced by these SMEFT coefficients can therefore be compensated by shifts in the PDFs, leading to wider allowed intervals. This degeneracy is weaker for operators involving heavier quark flavours, whose luminosities are suppressed in the same kinematic region.

The same SMEFT-PDF interplay can be viewed from the PDF side by studying the posterior PDF luminosity uncertainties after profiling over SMEFT contributions. In this comparison, the quark flavour-general LFU scenario has a larger effect than the flavour-universal $(W,Y)$ case. For some luminosities, the broadening of the posterior PDF uncertainty is roughly twice as large in the LFU fit. This provides a complementary indication that allowing a larger set of quark-flavour structures opens additional directions in which SMEFT effects and PDF variations can partially mimic one another.

We also saw that angular information provides an additional way to reduce degeneracies in the fits. In one-parameter fits with PDFs fixed, merging the two angular bins has only a limited effect on the bounds, since much of the sensitivity already comes from the invariant-mass spectrum. Its impact becomes more important once several operator structures can contribute with similar mass dependence. In such cases, the angular dependence helps separate different chiral contributions and can remove degeneracies that are otherwise present after angular integration. We also found that angular information is especially relevant when PDFs are profiled, because the angular information gives the fit an additional handle to distinguish SMEFT effects from allowed PDF deformations.

These results point to some natural directions for future work.
Other high-mass Drell-Yan datasets should be implemented, both to obtain current constraints as well as to derive future sensitivity projections. These include other ATLAS and CMS dilepton datasets, as well as the inclusion of the charged-current single-lepton plus missing energy searches. The latter is particularly useful to test the operators with $SU(2)_L$-triplet structure.
The analysis can also be extended to all semileptonic operators, thus including scalar and tensor ones, as well as those that induce quark flavour-changing neutral currents.
Furthermore, sensitivity projections for the HL-LHC can be improved by implementing finer invariant-mass and angular binning than the setup used here, which is driven by the current CMS analysis.

\section*{Acknowledgments}

The authors thank M. Ubiali for facilitating access to the baseline PDF set used in Ref.~\cite{Greljo:2021kvv}.

\appendix

\section{Parton distribution functions}
\label{app:pdf}

\subsection{Prior PDF set and Hessian representation}
\label{app:pdf_prior}

The interpretation of high-mass Drell-Yan data in terms of SMEFT operators depends on the parton luminosities entering the corresponding quark-antiquark initial states.  PDF modelling therefore constitutes an important component of the uncertainty, and can affect the resulting bounds on flavour-dependent Wilson coefficients.

\begin{figure}[t]
    \centering   
    \includegraphics[width=0.65\linewidth]{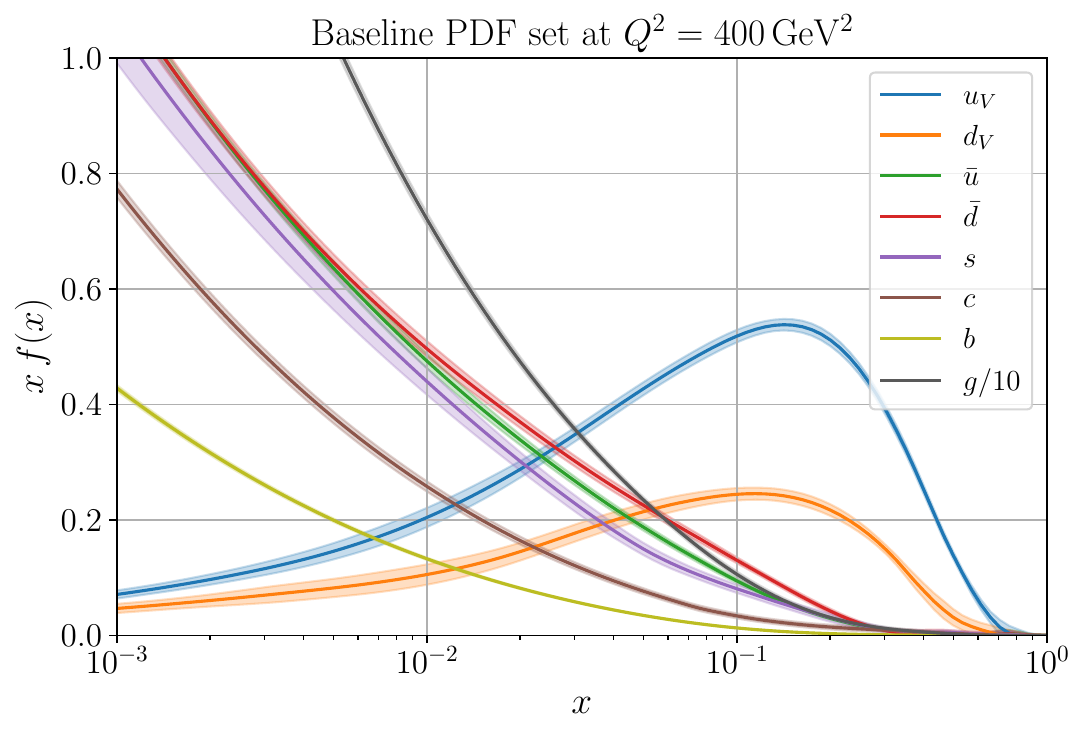}
    \caption{Baseline PDF set without high-mass Drell-Yan distributions for different channels at $Q^2 =400 \,  \mathrm{GeV}^2$.}
    \label{fig:pdf_summary}
\end{figure}

In this work we adopt the `DIS+DY (no HM)' PDF baseline of Ref.~\cite{Greljo:2021kvv}. This set is determined from DIS and Drell-Yan data, while excluding high-mass Drell-Yan distributions. This choice avoids using, as prior PDF information, the same high-energy tails that enter the SMEFT likelihood, and therefore provides a suitable setup to study how PDF uncertainties impact flavourful Drell-Yan constraints. A summary plot of the PDF set used in this study is shown in Figure~\ref{fig:pdf_summary}.

The `DIS+DY (no HM)' PDF set is originally provided as a Monte Carlo replica ensemble. For the profiling analysis we convert this ensemble into a Hessian representation. This rewrites the covariance of the replica ensemble in terms of independent directions that can be incorporated in the likelihood as nuisance parameters. We follow the MC2Hessian construction of Refs.~\cite{Carrazza:2015aoa,Carrazza:2016htc} and make use of LHAPDF~\cite{Buckley:2014ana} functionalities.

Let $f^{(r)}_{p}(x,Q)$ denote the $r$-th Monte Carlo replica of the $p$-th parton inside the list $\{u,d,s,c,b,\bar u,\bar d,\bar s,\bar c,\bar b\}_p$, with $r=1,\ldots,N_{\rm rep}$, and let $f^{(0)}_{p}(x,Q)$ be the central PDF, defined as the mean of the replica ensemble. The replicas are evaluated on a grid in momentum fraction and flavour at a reference scale $Q_0$. We then define the sampling matrix
\begin{equation}
X_{ar}
=
f^{(r)}_a(Q_0)-f^{(0)}_a(Q_0)~,
\label{eq:pdf_rep_samp}
\end{equation}
where the collective index $a=(p,n)$ denotes the parton flavour index $p$ and the $x$-grid point $x_n$. Equivalently, $a$ may be viewed as the flattened index $a=N_x(p-1)+n$, where $N_x$ is the number of grid points (on a sufficiently fine grid to sample the PDF correctly). The corresponding Monte Carlo covariance matrix in the discretised PDF space is
\begin{equation}
\mathrm{cov}_{ab}
=
\frac{1}{N_{\rm rep}-1}
\sum_{r=1}^{N_{\rm rep}}
X_{ar}X_{br}~.
\label{eq:pdf_mc_cov}
\end{equation}

The Hessian basis is obtained by diagonalising the covariance matrix. Equivalently, one may perform a singular-value decomposition of the sampling matrix,
\begin{equation}
X = U S V^T ~,
\label{eq:pdf_svd}
\end{equation}
where the columns of $U$ define orthonormal directions in the discretised PDF space and the singular values $s_\alpha$ determine the variance along each direction. The $\alpha$-th Hessian displacement is
\begin{equation}
\delta f^{(\alpha)}_a(Q_0)
=
\frac{s_\alpha}{\sqrt{N_{\rm rep}-1}}\,U_{a\alpha}~,
\label{eq:pdf_hessian_disp}
\end{equation}
so that the corresponding Hessian error member is
\begin{equation}
f^{(\alpha)}_a(Q_0)
=
f^{(0)}_a(Q_0)
+
\delta f^{(\alpha)}_a(Q_0)~.
\label{eq:pdf_hessian_member}
\end{equation} 
With this normalisation, a unit displacement along a Hessian direction corresponds to a one-standard-deviation PDF variation. In this way, retaining all non-zero singular directions reproduces the Monte Carlo covariance, while keeping only the first $N_{\rm eig}$ directions, ordered by decreasing singular value, gives a compressed Hessian set that preserves the dominant PDF uncertainty modes.

The treatment of PDF uncertainties through a Hessian error set, whose directions can be incorporated in the likelihood as nuisance parameters, has also been used for instance in Ref.~\cite{Torre:2020aiz}. We note, however, that the PDF basis adopted there is based on PDF4LHC15, which inherits high-mass Drell-Yan information through its constituent global fits (in particular from 7 TeV high-mass Drell-Yan single- and double-differential distributions from ATLAS and CMS).

\subsection{PDF posteriors}
\label{app:pdf_post}

\begin{figure}[t]
    \centering
    \includegraphics[width=0.95\linewidth]{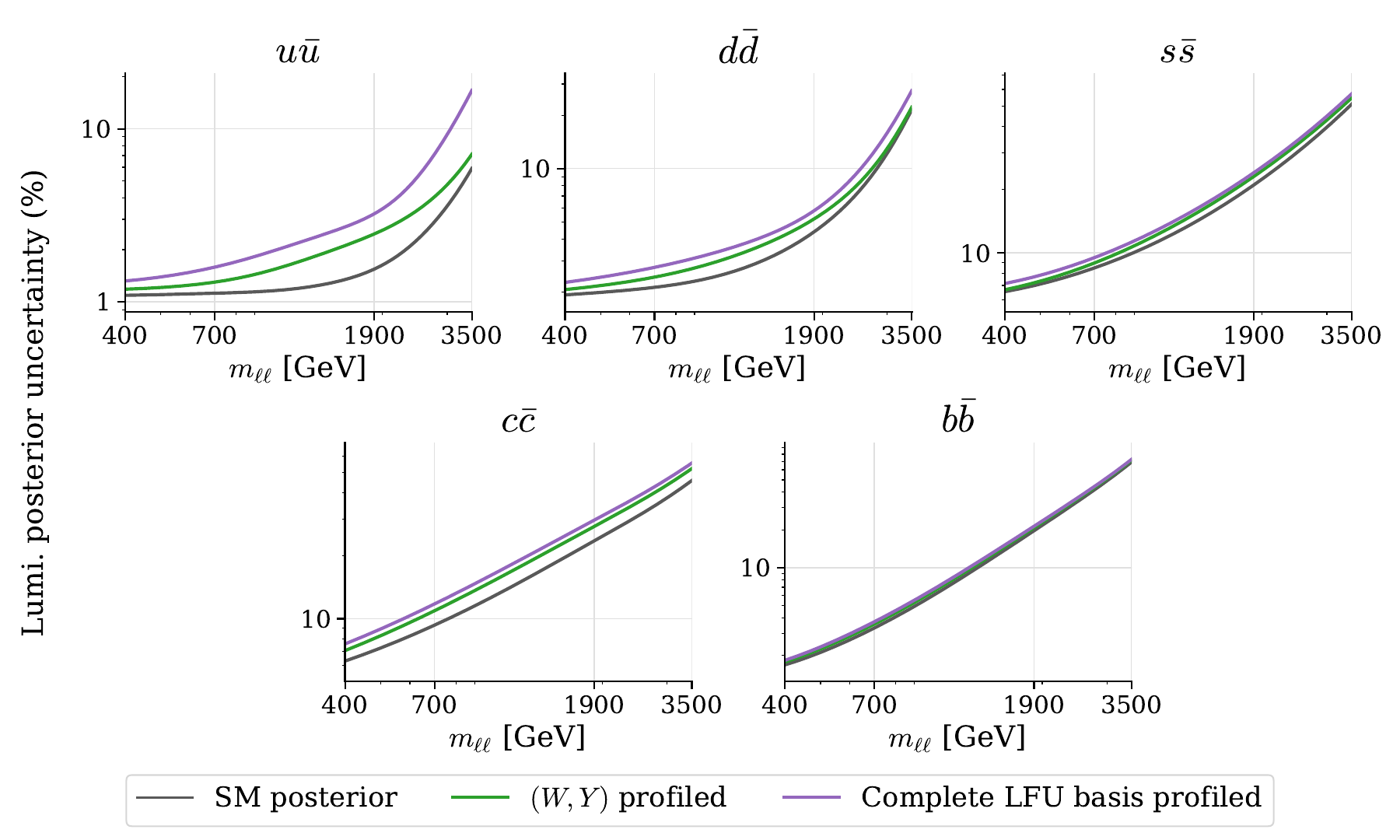}
    \caption{Luminosity posterior bands for different parton luminosities at the HL-LHC with respect to the SM PDF fit.}
    \label{fig:hllhc_lfu_wy_lumi_bands}
\end{figure}

In Figure~\ref{fig:hllhc_lfu_wy_lumi_bands} we show the corresponding luminosity posterior bands, normalised to the SM posterior fit. This information supplements the broadening factors shown in Figure~\ref{fig:hllhc_lfu_lumi_broad} where, as discussed in the main text, the sizeable broadening of the luminosity uncertainties induced by the EFT profiling occurs while the SM posterior uncertainties remain under control. 

\begin{figure}[t]
    \centering
    \includegraphics[width=0.95\linewidth]{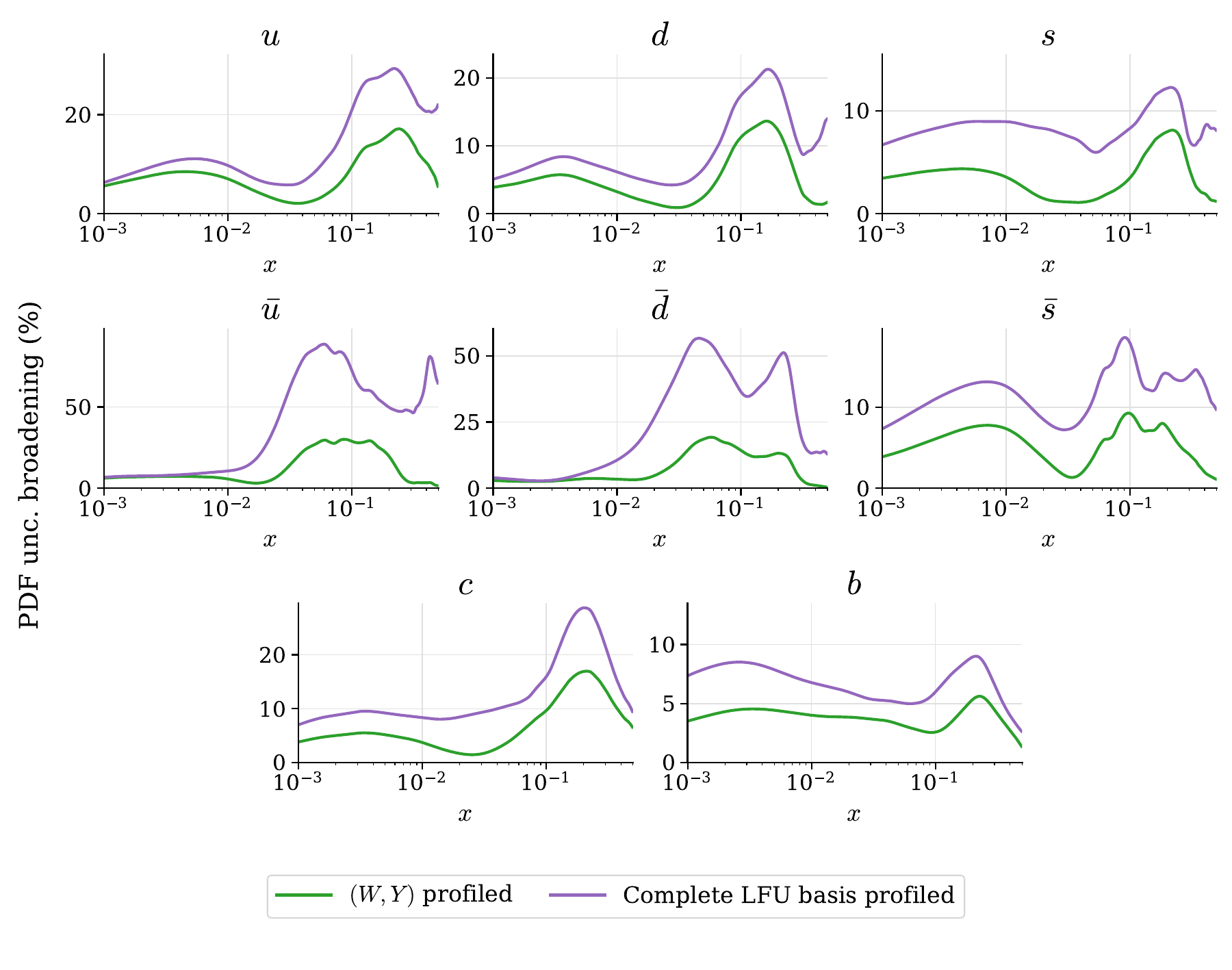}
    \caption{PDF uncertainty broadening with respect to the SM posterior for different flavours at the HL-LHC.}
    \label{fig:hllhc_wy_lfu_pdf_broad}
\end{figure}

Figure~\ref{fig:hllhc_wy_lfu_pdf_broad} shows the corresponding uncertainty broadening at the level of individual PDFs. As in the luminosity comparison, profiling over the complete LFU basis leads to a larger degradation than in the universal $(W, Y)$ case. A notable feature is that, within the light-quark channels most relevant for (neutral current) Drell-Yan, the broadening is more pronounced for the antiquark PDFs, in particular $\bar{u}$ and $\bar{d}$. A similar effect was observed in Ref.~\cite{Greljo:2021kvv}, although the effect is amplified here by the use of HL-LHC projections and the LFU basis. 
%

\section{Hadronic cross section}
\label{app:xsec}

In this appendix we collect the leading order relation between the partonic dilepton cross section (where EFT effects enter) and the hadronic prediction (where PDFs enter) differential in the dilepton invariant mass and in the Collins-Soper angle. This is the kinematic setup used for the theory predictions entering the two-dimensional bins in $m_{\ell \ell}$ and $\cos \theta^*$.

\subsection{Kinematics and the Collins-Soper angle}

Consider the partonic process
\begin{equation}
  q_i(p_1)\,\bar q_i(p_2)\to \ell_k^-(p_3)\,\ell_k^+(p_4)~.
\label{eq:partonic}
\end{equation}
The CMS analysis~\cite{CMS:2021ctt} uses the signed Collins-Soper angle of the negatively charged lepton:
\begin{equation}
 c\equiv \cos\theta^*
 =
 \frac{p_{z,\ell\ell}}{|p_{z,\ell\ell}|}
 \frac{2\left(p_3^+p_4^- - p_3^-p_4^+\right)}
 {m_{\ell\ell}\sqrt{m_{\ell\ell}^2+p_{T,\ell\ell}^2}},
 \qquad
 p_j^\pm=\frac{E_j\pm p_{z,j}}{\sqrt{2}}~,
\label{eq:cs-cms-def}
\end{equation}
where $j=3,4$, and $p_{\ell\ell}=p_3+p_4$. At leading order in the massless limit, Eq.~(\ref{eq:cs-cms-def}) reduces to
\begin{equation}
  c
  =
  \operatorname{sgn}(Y)\,
  \tanh\!\left(\frac{y_3-y_4}{2}\right),
  \qquad
  Y=\frac{y_3+y_4}{2}~.
\label{eq:cs-lo-rapidity}
\end{equation}
Here $Y$ is the dilepton rapidity. The LO momentum fractions are
\begin{equation}
  x_{1,2}=\frac{m_{\ell \ell}}{\sqrt{s}}e^{\pm Y}~,
\label{eq:x1x2}
\end{equation}
so that $\operatorname{sgn}(Y)=\operatorname{sgn}(x_1-x_2)$. 

It is useful to distinguish $c$ from the partonic scattering angle
\begin{equation}
  \hat c\equiv \cos\theta~,
\end{equation}
where $\theta$ is measured between the outgoing $\ell^-$, with momentum $p_3$, and the incoming quark direction in the partonic centre-of-mass frame. For the subprocess in Eq.~\eqref{eq:partonic},
\begin{equation}
  \hat t=(p_1-p_3)^2
  =
  -\frac{m^2}{2}(1-\hat c),
  \qquad
  \hat u=(p_1-p_4)^2
  =
  -\frac{m^2}{2}(1+\hat c)~.
\label{eq:mandelstam-chat}
\end{equation}

When embedding the partonic process in the $pp$ collision, there are two possible beam assignments. For
\begin{equation}
  q_i(x_1P_1)\,\bar q_i(x_2P_2)\to \ell_k^-(p_3)\ell_k^+(p_4)~,
\end{equation}
the incoming quark is aligned with beam 1, and
\begin{equation}
  \hat c=\operatorname{sgn}(Y)c~.
\label{eq:chat-ordered}
\end{equation}
For the reverse beam assignment,
\begin{equation}
  \bar q_i(x_1P_1)\,q_i(x_2P_2)\to \ell_k^-(p_3)\ell_k^+(p_4)~,
\end{equation}
the incoming quark is aligned with beam 2, and the partonic angle is reversed,
\begin{equation}
  \hat c=-\operatorname{sgn}(Y)c~.
\label{eq:chat-reverse}
\end{equation}
This sign flip is the origin of the interchange of the angular weights in the hadronic formula below.

\subsection{Partonic and hadronic cross sections}

We first recall the partonic expression of Eq.~\eqref{eq:partonic_xsec} introduced in Sec.~\ref{sec:SMEFT}. For the ordered subprocess $q_i(p_1)\bar q_i(p_2)\to \ell_k^-(p_3)\ell_k^+(p_4)$, the LO spin- and colour-averaged differential cross section can be written as
\begin{equation}
\frac{d\hat\sigma^{q_i \ell_k}}{d\hat t}
=
\frac{1}{48\pi \hat s^2}
\left[
\hat u^2 A_{q_i \ell_k}(\hat s)
+
\hat t^2 B_{q_i \ell_k}(\hat s)
\right]~,
\label{eq:partonic-xsec-ab}
\end{equation}
where
\begin{align}
A_{q_i \ell_k}(\hat s)
&\equiv
\left|F^{q_i \ell_k}_{q_{L}\ell_{L}}(\hat s)\right|^2
+
\left|F^{q_i \ell_k}_{q_{R}\ell_{R}}(\hat s)\right|^2~,
\label{eq:Aik-def}
\\
B_{q_i \ell_k}(\hat s)
&\equiv
\left|F^{q_i \ell_k}_{q_{L}\ell_{R}}(\hat s)\right|^2
+
\left|F^{q_i \ell_k}_{q_{R}\ell_{L}}(\hat s)\right|^2~.
\label{eq:Bik-def}
\end{align}
At LO, $\hat s=m^2$. Substituting Eq.~\eqref{eq:mandelstam-chat} into Eq.~\eqref{eq:partonic-xsec-ab} gives
\begin{equation}
\frac{d\hat\sigma^{q_i \ell_k}}{d\hat t}
=
\frac{1}{192\pi}
\left[
(1+\hat c)^2 A_{q_i \ell_k}(m^2)
+
(1-\hat c)^2 B_{q_i \ell_k}(m^2)
\right] ~.
\label{eq:partonic-xsec-chat}
\end{equation}
Convolving this with the PDFs gives the LO factorised hadronic cross section
\begin{equation}
  d\sigma^{\ell_k}
  = \sum_{q_i}
  f_{q_i}(x_1,\mu_F)\,
  f_{\bar q_i}(x_2,\mu_F)\,
  dx_1\,dx_2\,
  \frac{d\hat\sigma^{q_i \ell_k}}{d\hat t}\,d\hat t~+(x_1 \leftrightarrow x_2) ~,
\end{equation}
where $\mu_F$ is the factorisation scale.
Changing variables from $(x_1,x_2)$ to $(m,Y)$ gives
\begin{equation}
  dx_1\,dx_2=\frac{2m}{s}\,dm\,dY,
  \qquad
  Y_{\max}=\ln\!\left(\frac{\sqrt{s}}{m}\right).
\label{eq:x-jacobian}
\end{equation}
For fixed $m$, the change from $\hat t$ to the measured Collins-Soper variable gives
\begin{equation}
  \left|\frac{\partial \hat t}{\partial c}\right|=\frac{m^2}{2}~.
\label{eq:t-c-jacobian}
\end{equation}
Using Eqs.~\eqref{eq:chat-ordered} and \eqref{eq:chat-reverse}, the hadronic cross section for a given charged-lepton flavour $k$ is
\begin{equation}
\begin{aligned}
\frac{d^2\sigma^{\ell_k}}{dm\,dc}
&=
\frac{m^3}{192\pi s}
\sum_i
\int_{-Y_{\max}}^{Y_{\max}} dY
\Bigg\{
f_{q_i}(x_1,\mu_F)\,f_{\bar q_i}(x_2,\mu_F)
\\
&\quad\times
\left[
\bigl(1+\operatorname{sgn}(Y)c\bigr)^2 A_{q_i \ell_k}(m^2)
+
\bigl(1-\operatorname{sgn}(Y)c\bigr)^2 B_{q_i \ell_k}(m^2)
\right]
\\
&\quad+
f_{\bar q_i}(x_1,\mu_F)\,f_{q_i}(x_2,\mu_F)
\\
&\quad\times
\left[
\bigl(1-\operatorname{sgn}(Y)c\bigr)^2 A_{q_i \ell_k}(m^2)
+
\bigl(1+\operatorname{sgn}(Y)c\bigr)^2 B_{q_i \ell_k}(m^2)
\right]
\Bigg\}~.
\end{aligned}
\label{eq:hadronic-fullY}
\end{equation}
In the second term, with reversed $x_1,x_2$, the weights multiplying $A_{q_i \ell_k}$ and $B_{q_i \ell_k}$ are exchanged because $\hat c$ is defined with respect to the incoming quark direction.

Finally, the expected cross section for a given dilepton channel $\ell_k^+ \ell_k^-$, in a bin of invariant mass $\left[m_{\min} , m_{\max} \right]$ and Collins-Soper angle $\left[c_{\min} , c_{\max} \right]$, is
\begin{equation}
  \sigma^{\ell_k}_{\rm bin}
  =
  \int_{m_{\min}}^{m_{\max}} dm
  \int_{c_{\min}}^{c_{\max}} dc\,
  \frac{d^2\sigma^{\ell_k}}{dm\,dc}~.
\label{eq:bin-xsec}
\end{equation}

\begin{figure}[t]
    \centering   
    \includegraphics[width=0.66\linewidth]{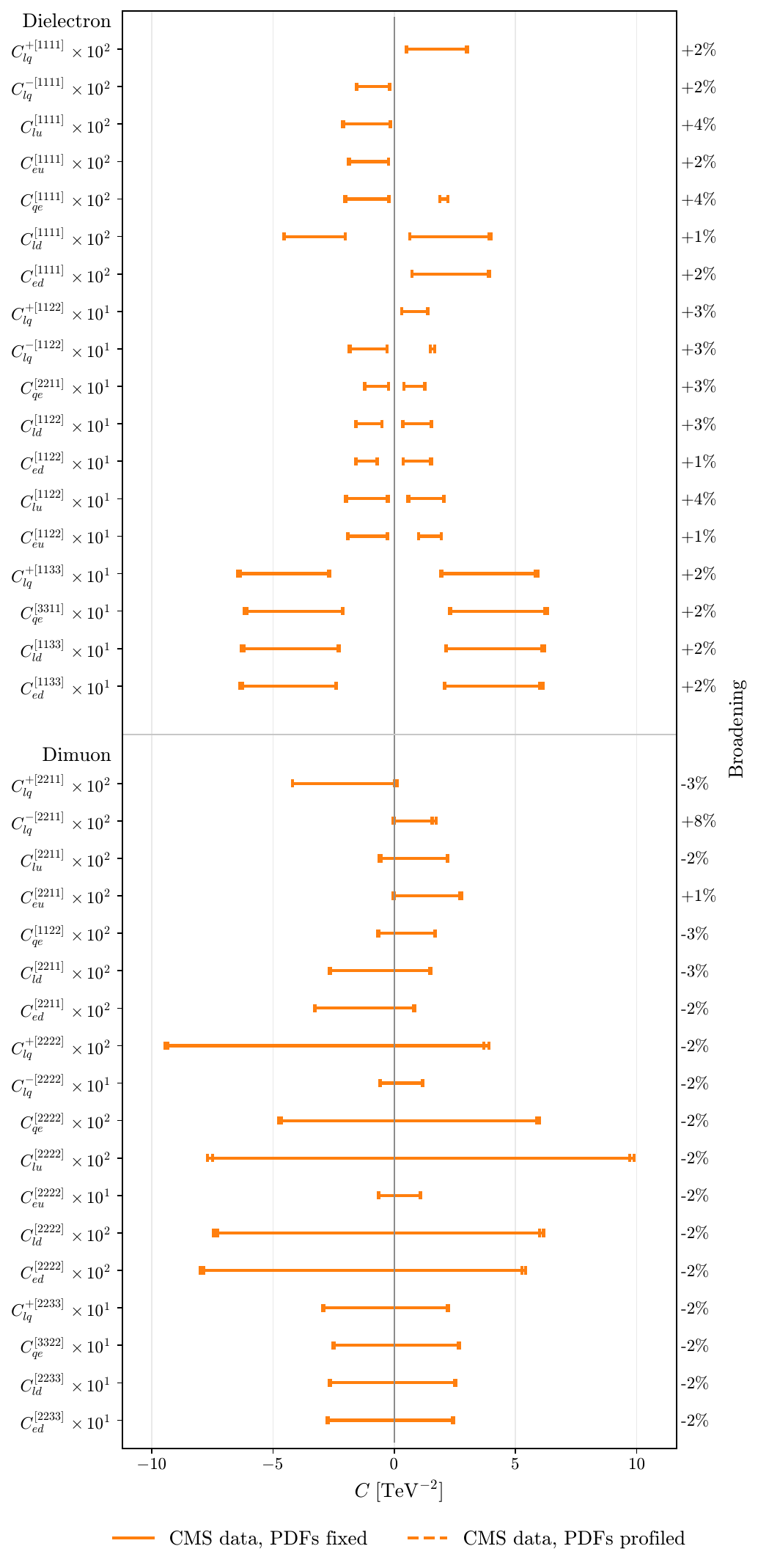}
    \caption{One-parameter fits for both lepton channels with current data.}
    \label{fig:eft_summary_plot}
\end{figure}
\begin{figure}[t]
    \centering   
    \includegraphics[width=0.66\linewidth]{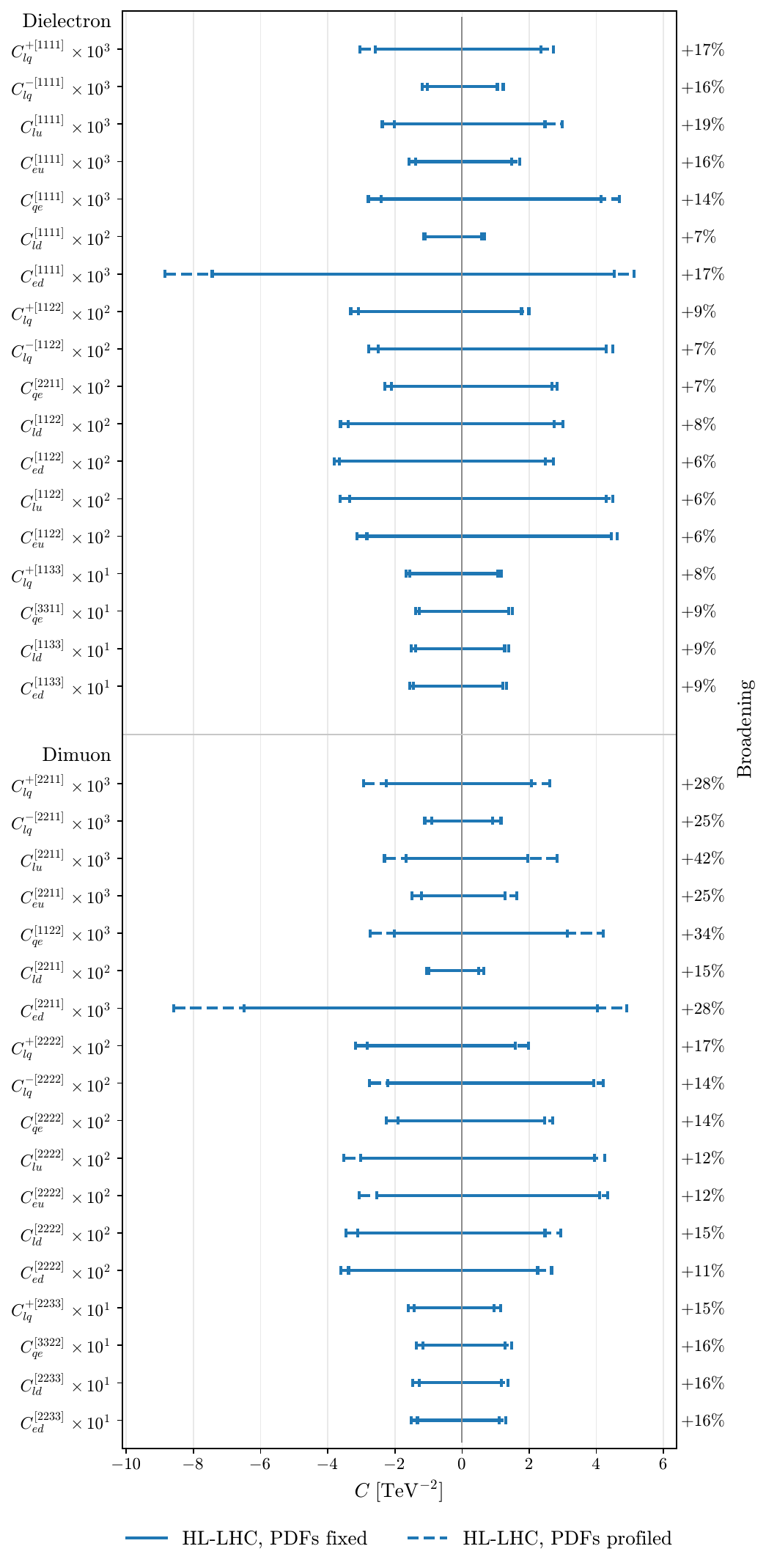}
    \caption{One-parameter fits for both lepton channels at the HL-LHC.}
    \label{fig:eft_summary_plot_hllhc}
\end{figure}
%

\section{Flavour-general results}
\label{app:fl_general}

If the LFU assumption is lifted, the broadening effect of PDF uncertainties in SMEFT constraints is expected to be smaller. This is because PDF variations are always universal in lepton flavour, so a non-LFU SMEFT direction is not aligned to PDF variations.
This could also be understood in terms of the LFU ratio of the cross sections in the two lepton flavours, introduced in Ref.~\cite{Greljo:2017vvb}. This observable is not sensitive at all to PDF effects, at least in the SM limit, while it is sensitive to non-LFU SMEFT effects.

Figure~\ref{fig:eft_summary_plot} summarises the one-parameter constraints obtained from the current data in the full flavour-general SMEFT basis. The upper block corresponds to the dielectron channel, while the lower block corresponds to the dimuon channel. For each coefficient, we report the $95 \%$ CL interval with fixed PDFs and after PDF profiling. In order to compute the broadening factor for the disconnected regions that appear in the dielectron channel, the components are matched separately and the quoted value is the smaller relative broadening among the matched components.

The impact of the PDF profiling on the current data bounds is generally mild in the fully flavour-general fit. In the dimuon channel, all intervals remain compatible with the SM point. The largest change in the interval width is observed for $C_{lq}^{-[2211]}$, at the level of about $8\%$, while most other coefficients change only by a few percent. Some dimuon intervals are slightly narrower after profiling. This happens because, while profiling always lowers the $\chi^2$ curve at a fixed EFT coefficient value, the quoted intervals are defined from $\Delta\chi^2$ with respect to the profiled minimum, which also gets lower. It can happen that the $\chi^2$ at the best-fit point lowers more than the $\chi^2$ at the interval boundaries, in which case one sees a restriction of the 95\% CL interval.

The dielectron channel shows a qualitatively different behaviour. The bounds are displaced from the SM point, and in several cases split into disconnected components. This reflects the tension already present in the dielectron data with respect to the SM prediction. As already seen in the dimuon case, the PDF-induced changes remain small, typically at the level of $1$--$4\%$.

Figure~\ref{fig:eft_summary_plot_hllhc} summarises the constraints obtained from the HL-LHC projections in this more flavour-general basis.
The effect of PDF profiling is visibly larger than for current data, as expected from the reduced experimental uncertainties in the HL-LHC projection. The largest broadenings occur in the dimuon channel for first-generation quark operators, reaching about $40\%$ for $C_{lu}^{[2211]}$ and typically $25$--$35\%$ for the other most affected first-generation directions.  The dielectron channel is less affected, with a maximum broadening of about $20\%$.
Second- and third-generation quark operators show milder changes, generally at the level of a few tens of percent or less.  Overall, the PDF effect in this flavour-general basis remains significantly smaller than in the LFU scenario, where several first-generation directions broaden by order one and many heavier-flavour directions still receive $30$--$60\%$ corrections.

\bibliographystyle{JHEP}
\bibliography{biblio}

\end{document}